\documentclass[]{aa}
\pdfoutput=1
\usepackage{amsmath}
\usepackage{txfonts} 
\usepackage{graphicx}
\usepackage{subfigure}
\usepackage{color}
\usepackage{bm}
\usepackage{natbib}
\usepackage{url}
\usepackage[center, labelfont=bf]{caption}
\usepackage[version=3]{mhchem} 

\usepackage{tabularx}
\usepackage{txfonts}
\usepackage{longtable}
\usepackage{hhline}
\usepackage{arydshln}
\usepackage{multirow}
\usepackage{multicol}
\usepackage{lscape}
\usepackage{array}
\usepackage{rotating}

\newcommand{\degrees}{$\degr$C}
\newcommand{\vol}{\textit{vol}\%}
\newcommand{\mass}{\textit{wt}\%}

\newlength{\pointwidth}

\newcommand{\co}{CO}
\newcommand{\cod}{CO$_{2}$}
\newcommand{\sic}{SiC}
\newcommand{\hydrogen}{H$_{2}$}
\newcommand{\methane}{CH$_{4}$}
\newcommand{\ethane}{C$_{2}$H$_{6}$}
\newcommand{\ethilene}{C$_{2}$H$_{4}$}
\newcommand{\methanol}{CH$_{3}$OH}

\newcommand{\ethanol}{CH$_{3}$CH$_{2}$OH}

\newcommand{\propanol}{CH$_{3}$CH$_{2}$CH$_{2}$OH}

\newcommand{\butanol}{CH$_{3}$CH$_{2}$CH$_{2}$CH$_{2}$OH}

\newcommand{\formaldehyde}{H$_{2}$CO}
\newcommand{\acetone}{(CH$_{3}$)$_{2}$CO}

\newcommand{\oxygent}{O$_{2}$}
\newcommand{\ammonia}{NH$_{3}$}
\newcommand{\water}{H$_{2}$O}
\newcommand{\nitrogen}{N$_{2}$}
\newcommand{\acetaldehyde}{CH$_{3}$CHO}
\newcommand{\formamide}{NH$_{2}$CHO}
\newcommand{\argon}{Ar}

\begin{document}

\title{Study of Fischer-Tropsch Type reactions on chondritic meteorites}

  \author{V. Cabedo,  
  	       \inst{1,3},
	      J. Llorca,
  	       \inst{2}
  	      J.M. Trigo-Rodriguez,
  	       \inst{3,4}
  	      A. Rimola,
  	        \inst{5}
  	       }

\institute{
Astrophysics department, CEA/DRF/IRFU/DAp, Universit\'{e} Paris Saclay, UMR AIM, F-91191 Gif-sur-Yvette, France \\
\email{victoria.cabedo@cea.fr} 
\and
Institut de Tècniques Energètiques and Departament d'Enginyeria Química, Universitat Politècnica de Catalunya, Barcelona, Catalonia, Spain
\and
Institute of Space Sciences (CSIC), Meteorites, Minor Bodies and Planetary Sciences Group, Campus UAB, Carrer de Can Magrans, s/n, 08193, Barcelona, Catalonia, Spain
\and
Institut d'Estudis Espacials de Catalunya (IEEC), Gran Capità, 2 - baix, 08034, Barcelona, Catalonia, Spain
\and
Departament de Química, Universitat Autònoma de Barcelona, 08193 Bellaterra, Catalonia, Spain}

\abstract
{How simple organic matter appeared on Earth and the processes by which it transformed into more evolved organic compounds, which ultimately led to the emergence of life, is still an open topic. Different scenarios have been proposed, the main one assumes that simple organic compounds were synthesized, either in the gas phase or on the surfaces of dust grains, during the process of star formation, and were incorporated into larger bodies in the protoplanetary disk. Transformation of these simple organic compounds in more complex forms is still a matter of debate. Recent discoveries point out to catalytic properties of dust grains present in the early stellar envelope, which can nowadays be found in the form of chondrites. The huge infall of chondritic meteorites during the early periods of Earth suggests that the same reactions could have taken place in certain environments of the Earth surface, with conditions more favorable for organic synthesis.}
{This work attempts the synthesis of simple organic molecules, such as hydrocarbons and alcohols, via Fischer-Tropsch Type reactions supported by different chondritic materials under early-Earth conditions, to investigate if organic synthesis can likely occur in this environment and which are the differences in selectivity when using different types of chondrites.}
{Fischer-Tropsch-type reactions are investigated from mixtures of \co\ and \hydrogen\ at 1 atm of pressure on the surfaces of different chondritic samples. The different products obtained are analyzed \textit{in situ} by gas chromatography.}
{Different Fischer-Tropsch reaction products are obtained in quantitative amounts. The formation of alkanes and alkenes being the main processes. Formation of alcohols also takes place in a smaller amount. Other secondary products were obtained in a qualitative way.}
{Chondritic material surfaces have been proven as good supports for the occurrence of organic synthesis. Under certain circumstances during the formation of Earth, they could have produced a suitable environment for these reactions to happen.}


\authorrunning{V. Cabedo et. al.}
\titlerunning{FTT reactions in chondritic meteorites}
\maketitle


\section{Introduction} 

Life on Earth likely arose from a series of particular conditions, which remain to be unambiguously determined. It has been envisioned that the early Earth was exposed to different sources of energy, such as light, heat and reduction-oxidation (redox) potentials. Such environment also received a large flux of extraterrestrial material,  at least until the end of the Late Heavy Bombardment (LHB, \citealt{Bottke2012}).  At some point, the conditions in the atmosphere and surface of our planet might have promoted an increase in molecular complexity (\citealt{Rotelli2016}) which, later on, allowed the recombination of organic molecules to form more complex, interacting and self-replicating systems (\citealt{Walde2005}, \citealt{Schulze-Makuch2008}).

The synthesis of interstellar complex organic molecules (iCOMs) in the solar nebula and during the process of star formation has been extensively studied during the last years (\citealt{Herbst2017}). The simplest organic compounds, such as formaldehyde (\formaldehyde) and methanol (\methanol), and other iCOMS, for instance acetaldehyde (\acetaldehyde) and formamide (\formamide), have been repeatedly observed from starless cores to star forming regions (\citealt{Vasyunina2014}, \citealt{Lopez-Sepulcre2015}, \citealt{Bianchi2018}, \citealt{McGuire2018}, \citealt{vanGelder2020}). Robust evidences indicate that the formation of iCOMs takes place on the surface of interstellar grains (\citealt{Kress2001}, \citealt{Sekine2006}, \citealt{Linnartz2015}, \citealt{Oberg2016},  \citealt{JER2019}, \citealt{Zamirri2019}, \citealt{Martin-Domenech2020}), which with time will  coagulate and form the embryos of rocky planets and minor bodies. 

Chondrites are a class of meteorites that come from undifferentiated bodies (they never melted), such as small asteroids or comets, and that have suffered very low chemical or thermal alteration processes (\citealt{Zolensky2008}). Chondrites come from the direct accretion of the primordial materials forming the protoplanetary disk, so they are believed to be the most pristine samples of the dust grains forming our early Solar System (\citealt{Weisberg2006}, \citealt{Trigo2015}) preserving their primordial components: metal grains, sulfides and igneous silicate-rich spherules called chondrules (hence the name chondrites). Particularly interesting are carbonaceous chondrites (CCs), since they contain about 1\--4 \% of carbon in mass (\citealt{Remusat2007}, \citealt{Alexander2017}), which is found in the form of organic matter aggregates (\citealt{Trigo2019}). This organic matter is thought to have been synthesized in the solar nebula, some of these reactions being catalyzed by the surfaces of the metallic phase present in the dust (\citealt{Llorca2000}).

Surface reactions in solid-dry environments can play a central role in the synthesis of biomolecular precursors and other large complex organic molecules of extrasolar origin, which have been identified in undifferentiated meteorites (\citealt{Ehrenfreund2000}, \citealt{Septhon2002}, \citealt{Bada2002}). Indeed, different studies evidence that important prebiotic molecules can be formed via processing of interstellar ices (\citealt{Ciesla2012}, \citealt{Fedoseev2015}, \citealt{Chuang2015}, \citealt{Chuang2017}, \citealt{Fedoseev2017}), including amino acids (\citealt{Bernstein2002}, \citealt{Munoz-Caro2002}, \citealt{Elsila2007}, \citealt{Nuevo2008}, \citealt{Lee2009}). More recently, non-energetic ice chemistry (i.e., without the need of energetic external triggers processing the ice) has been demonstrated to be also operative in the formation of glycine (\citealt{Krasnokutski2020}, \citealt{Ioppolo2020}). However, investigations on the synthesis of prebiotic compounds on meteoritic minerals and/or analogues under dry conditions (i.e., through gas-surface reactions) is practically missing.

Since \hydrogen\ and \co\ are the most abundant gas-phase compounds in most of the astrophysical environments, alongside that metallic components usually used as catalytic supports are commonly found in meteoritic materials, gas-surface Fischer-Tropsch Type (FTT) reactions are here invoked as possible synthetic routes for part of the organic content (i.e., hydrocarbons and alcohols) identified in these asteroidal bodies (\citealt{Ferrante2000}, \citealt{Sekine2006}). The Fischer-Tropsch synthesis consists of a set of chemical reactions that produce hydrocarbons and fuels from mixtures of \co\ and \hydrogen\ (see Equations \ref{eq:parafines_formation} and \ref{eq:olefines_formation}). Additionally, it can also produce secondary reactions such as the formation of alcohols (Equation \ref{eq:alcohols_formation}) or the disproportion of \co\ (Equation \ref{eq:CO_disproportion}, \citealt{Schulz1999}, \citealt{Mahmoudi2017}). FTT processes proceed under the presence of a metal catalyst, generally Fe, Co, Ni and Ru. The mechanism of the reaction and the product selectivity and distribution vary depending on the metal used as support and on the reaction conditions (\citealt{Mahmoudi2017}). 

\begin{equation}
    n\ce{CO_{(g)}} + (2n+1)\ce{H2_{(g)}} \rightleftharpoons \ce{C_{n}}\ce{H_{2n+2(g)}} + n\ce{H2O_{(g)}}
    \label{eq:parafines_formation}
\end{equation}

\begin{equation}
    n\ce{CO_{(g)}} + (2n)\ce{H2_{(g)}} \rightleftharpoons \ce{C_{n}}\ce{H_{2n(g)}} + n\ce{H2O_{(g)}}
    \label{eq:olefines_formation}
\end{equation}

\begin{equation}
    n\ce{CO_{(g)}} + (2n)\ce{H2_{(g)}} \rightleftharpoons \ce{C_{n}}\ce{H_{2n+1}}\ce{OH_{(g)}} + (n-1)\ce{H2O_{(g)}}
    \label{eq:alcohols_formation}
\end{equation}

\begin{equation}
    2\ce{CO_{(g)}} \rightleftharpoons \ce{C_{(s)}} + \ce{CO2_{(g)}}
    \label{eq:CO_disproportion}
\end{equation}

The advent of organic chemical reactions in primordial grains of the solar nebula and the protoplanetary disk is of fundamental relevance, as these materials were lately processed to form the seeds of planets, comets and meteorites (\citealt{Brearley1998}). Moreover, the flux of chondritic material was important during the early periods of the Earth, which could provide the support and conditions necessary for these reactions to happen. The vapor clouds produced by the impacts of these bodies and the presence of these particular metal alloys, with the high pressure and high temperature conditions, could have established an adequate environment for complex chemistry to take place. 

If FTT reactions can be extrapolated to environments similar to the available conditions present in the primeval Earth, chondritic meteorites could be responsible not only of bringing large amounts of organic material and water to the protoplanet, but also of promoting reactions of paramount astrobiological relevance in the emergence of prebiotic chemistry. Therefore, the aim of this work is to explore this possibility, by simulating FTT processes, in an environment similar to the primeval atmosphere of Earth.

\section{Experimental procedure}

    \subsection{Sample description and preparation}
        
        \begin{table*}[h!]
            \centering
            \caption{Average petrologic characteristics of the chondrite groups used in this work. Adapted from \citet{Weisberg2006}.}
            \begin{tabular}{c c c c c c c}
            \hline
            \vspace{-5pt}
            &\\
            Type & Chondrule abund.* & Matrix abund. & CAI abund. & Metal abund. & Average chondrule diameter & Organic content \\
            & (\vol) & (\vol) & (\vol) & (\vol) & (mm) & (\mass)\\
            \hline
            &\\
            CR & 50-60 & 30-50 & 0.5 & 5-8 & 0.7 & 0.8-2.6  \\
            H & 60-80 & 10-15 & $\ll$ 1 & 8 & 0.3 & 0 \\
            &\\
            \hline
            \end{tabular}
            \begin{list}{}{}
                \centering
                    \item * Chondrule abundance includes lithic and mineral fragments
            \end{list}
            \label{table:petrologyDescrp}
        \end{table*}
    
        Two different meteoritic samples were used to prove their surface activity, provided by different collections and chosen by their availability. Samples of CCs and Ordinary Chondrites (OCs) have been chosen because of their primitive composition, with isotopic composition similar to solar, and their relative low degree of alteration. These characteristics make them good representatives of the dust present in the protoplanetary disk. Moreover, both classes cover an important percentage of the recovered meteorites that fall into Earth-surface, and therefore, by extrapolation, they are also good representatives of the kind of materials fallen during the early stages of Earth formation.
        
        KG 007 is an OC of subtype H and petrological type 6 (\citealt{MB_KG007}). Those are distinguished by presenting sub-solar Mg/Si and refractory/Si ratios, to exhibit oxygen isotope composition above the terrestrial fractionation line, and to contain a large volume percentage of chondrules, with only 10-15 \vol\ of fine-grained matrix (\citealt{Weisberg2006}). They do not contain organic matter in their matrix. The H (high-iron) group of OCs are characterized by their high siderophile element content, with metal abundance of around 8 \vol. Petrological type 6 designates chondrites that have suffered metamorphism under conditions sufficient to homogenize all mineral compositions, but melting did not occur and therefore there is no phase differentiation. KG 007 was obtained already grinded and in two different phases: one containing the full meteoritic composition (Sample B) and another one only containing the metallic inclusions separated by magnetic means (Sample C). 
        
        NWA 801 is a CC of subtype R and petrologic type 2 (\citealt{MB_NWA801}). CCs are very primitive materials that have Mg/Si ratios near to solar values and oxygen isotope compositions below the terrestrial fractionation line (\citealt{Weisberg2006}). The Renazzo subgroup (R) is distinguished by large and abundant porphyritic chondrules (50 \vol). It has few refractory inclusions and abundant metal percentage, between 5-8 \vol, and a fine-grained matrix, which is commonly hydrated occupying up to 50 \vol. As mentioned above, they are important because they contain up to 4 \%\ of organic matter aggregates in their matrix (\citealt{Remusat2007}, \citealt{Alexander2017}, \citealt{Trigo2019}). Petrological type 2 designates chondrites characterized by abundant hydrated minerals and abundant fine-grained matrix, with Ni-bearing sulfides. NWA 801 (Sample D) was obtained as a small chunk and was completely grinded before introducing it in the reactor, using a glass mortar until obtaining a very fine dust. Details on the petrology and composition of each chondrite type used in the samples can be found in Table \ref{table:petrologyDescrp}
        
        For each sample, 0.5 g of meteoritic dust has been mixed with 2.5 g of \sic. \sic\ is a very inert compound and its mixing assures a constant volume of the samples, so the contact time between the reactants and the samples is optimized to maximize the reaction rate, assure reproductibility and allow comparison between the samples. Details on the sample preparations can be found in Table \ref{table:sampleDescrp}. Sample A is used as the blank analysis in order to follow the progress of the reactions without meteoritical samples and to assure that there is no chemical activity coming from \sic. Sample E is used as second blank, with meteoritical content from NWA 801 but without passing \co\ and \hydrogen\ as reactants and circulating \argon\ instead, to ensure that the detected products are not coming from the meteorite itself due to the presence of organic matter in the matrix of CCs. This procedure is not done for KG 007 because it is an OC and therefore it does not contain any organic matter in the matrix.
        
        Due to the relatively rare nature of the samples, and the lack of enough material, we were unable to produce duplicates of each sample and of each experiment.

    \subsection{Fischer-Tropsch reactions}
    
        The reaction conditions were the same for all the samples. The total weight of the samples was introduced in a closed  stainless-steel reactor with an internal diameter of 6 mm, a total volume of 5 mL and heated by a furnace, where a flux of \co\ and \hydrogen\ was circulated. The conditions were selected to resemble those present on the primitive Earth surface:
        
        \begin{itemize}
            \item The ratio of reactants was set to \hydrogen:\co\ = 4, in consistency to the estimated primitive ratios of the primitive Earth atmosphere, i.e., \hydrogen:\co\ = 0.03 - 10 (\citealt{Kress2001}). The fluxes of the reactants were set to a total of 40 mL/min, with \hydrogen\ = 32 mL/min and \co\ = 8 mL/min. For Sample E, \argon\ was circulated as an inert gas with a flux of 40 mL/min.
            
            \item Pressure was set to 1 atm to simulate the average pressure on the Earth surface.
            
            \item Since relevant production of FTT reactions starts at 400 K (\citealt{Sekine2006}), here the temperature was increased manually from 200 to 600 \degrees\ in steps of 50 \degrees.
        \end{itemize}
        
\begin{figure}[h!]
            \centering
            \includegraphics[scale=0.4]{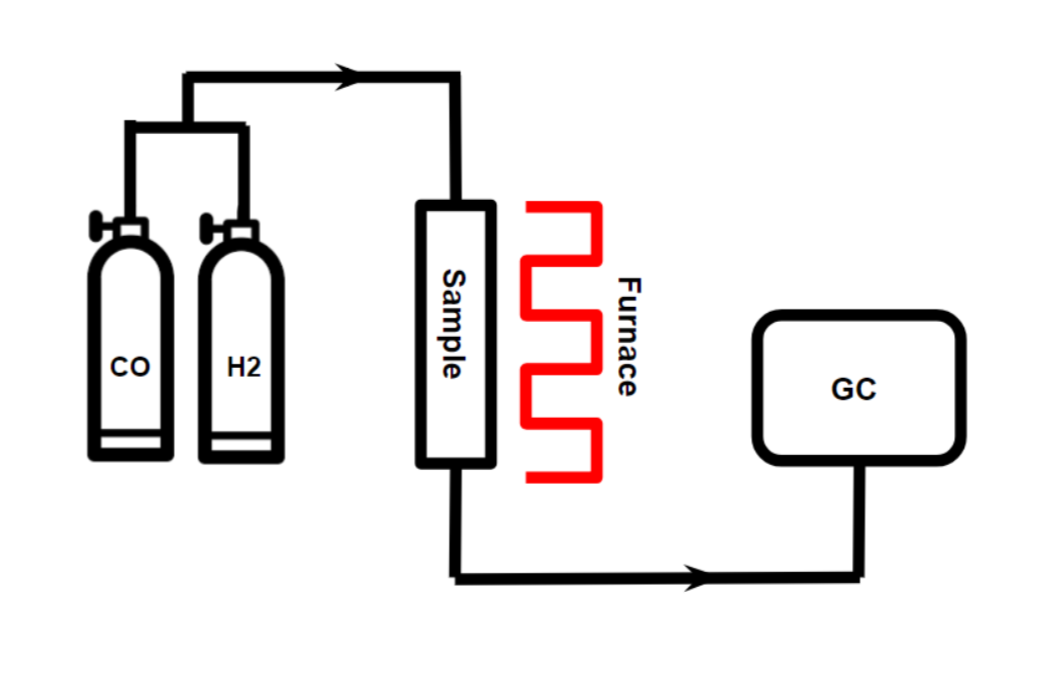}
            \caption{Schematic of the experimental setup. \co\ and \hydrogen\ are passed through the heated reactor containing the samples, and the resulting gas is injected to the GC.}
            \label{fig:experimental_setup}
        \end{figure}

        Products were analyzed \textit{on line} with an Agilent 3000A micro Gas Chromatography (GC) system. The GC is directly connected to the outlet of the reactor by stainless steel tubes. A schematic of the experimental setup is shown in Fig. \ref{fig:experimental_setup}. Three different columns were used in order to be able to analyze products with different polarities: an 8 m PLOTU for hydrocarbon analysis, a 10 m STABILWAX for oxygenated compound analysis, and a MS5A for non-polar compound analysis. Each product  was quantified with calibration standards and using the GC software \textit{Soprane}. Standards and reaction gases were supplied by Nippon Gases and were 99.999\%\ purity.
        
        Reactions were carried out during \textit{ca.} 3h, though the time varies depending on the stability of the production. Products are analyzed continuously, with injections to the GC every 5 min. After each experiment, the reactor was cleaned with a brush, using only water and then dried in an oven at 200 \degrees.
        
        \begin{table*}
        \centering
        \caption{Sample description}
        \begin{tabular}{c c c c c c c c c}
        \hline
        \vspace{-5pt}
        &\\
        Sample & Meteorite & Type & Subtype & Weight & \sic\ weight & Total weight & Gas flow  \\
        & & & & (g) & (g) & (g) & \\
        \hline
        &\\
        A & - & - & - & - & 3.0027 & 3.0027 & \hydrogen/\co \\
        B & KG007 & OC & H6 & 0.5012 & 2.4996 & 3.0008 & \hydrogen/\co \\
        C & KG007 & OC & H6 & 0.5158 & 2.4929 & 3.0087 & \hydrogen/\co \\
        & (magnetic phase) & & & & \\
        D & NWA 801 & CC & CR2 & 0.5256 & 2.4814 & 3.0070 & \hydrogen/\co \\
        E & NWA 801 & CC & CR2 & 0.5060 & 2.4986 & 3.0046 & \argon \\
        \hline
        \end{tabular}
        \begin{list}{}{}
            \centering
                \item Instrumental errors are $\pm$ 0.0001g 
        \end{list}
        \label{table:sampleDescrp}
        \end{table*}

\section{Results}

    In the present experiments, production of hydrocarbons has been successfully achieved, in addition to other minor products such as primarily alcohols and other oxygenated compounds. The percentage of products obtained has been computed as the percentage of \co\ that has been transformed in each different product, including the production of solid carbon, following Equation \ref{eq:molar_fract}.
    
    \begin{equation}
        \begin{split}
            \chi_{P} &= \frac{n_{P}}{n_{\co, R}} \times 100 \\
            &= \frac{n_{P}}{n_{CO}^{i} - n_{CO}^{f}} \times 100
        \end{split}
        \label{eq:molar_fract}
    \end{equation}
    
    where $\chi_{P}$ is the molar fraction of the formed product (in $\%$), n$_{P}$ is the obtained mols of the formed product, and n$_{\co, R}$ the mols of \co\ that have been consumed, which are computed as the difference between the initial \co\ mols (n$_{\co}^{i}$) and the final \co\ mols detected (n$_{\co}^{f}$). The obtained chromatograms for all the experiments are shown in Appendix A.
    
    \subsection{Hydrocarbon production}
    
        \begin{figure}[h!]
            \begin{tabular}{m{8.5cm}}\includegraphics[width=9cm]{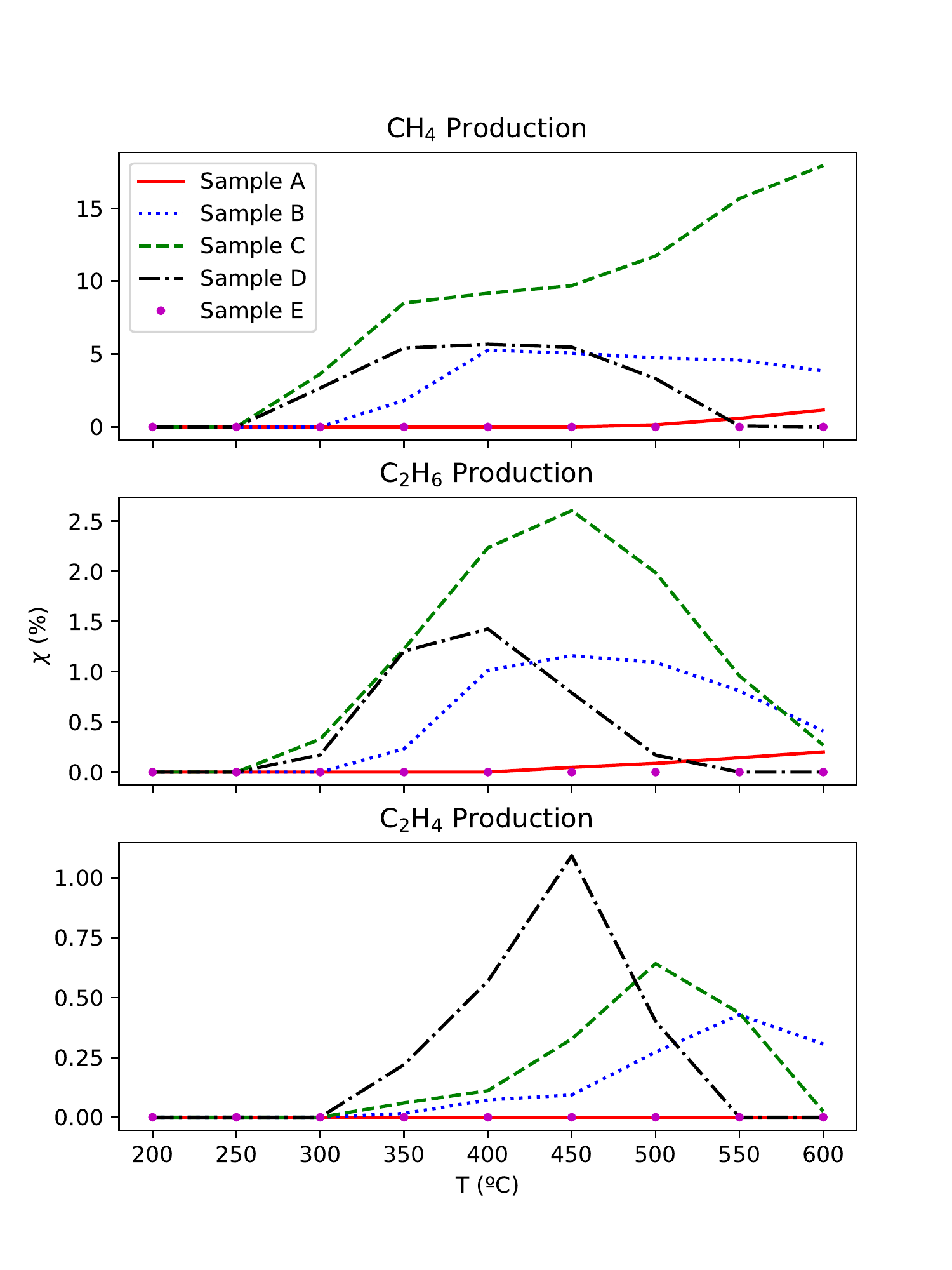} \\
            \end{tabular}
            \caption{Evolution of hydrocarbons production with temperature: \methane\ production (top), \ethane\ production (middle) and \ethilene\ production (bottom).}
            \label{fig:hydrocarbons_production}
        \end{figure}
    
        Figure \ref{fig:hydrocarbons_production} shows the evolution of the production of the different hydrocarbons with temperature. In the different experiments, alkanes have been successfully synthesized, methane (\methane) and ethane (\ethane) being the primary and the secondary products. 
        
        For samples B\--D, formation of \methane\ starts at 300 \degrees. For Samples B and C such a production still works at higher temperatures than 600 \degrees\, while for sample D it stops before reaching 550\degrees. 
       
        Production of \ethane\ also starts at 300 \degrees\, but it decreases at 550-600 \degrees\ for Sample B and C, while for sample D this happens at 400 \degrees. 
        
        Production of alkenes starts at a slightly higher temperature than than that for alkanes (i.e., 350\degrees). For sample B and C it starts to decrease when the temperature reaches 600\degrees\, while for sample D it stops before reaching 550\degrees. Formation of \ethilene\ starts to be detected when some amount of \ethane\ was produced and the maximum production of \ethilene\ occurs upon decrease of \ethane\, suggesting that both mechanisms are correlated, meaning that alkene formation starts when the amount of alkanes in the catalyst surface is large enough.
        
        Formation of alkynes is not detected, indicating that the energy barrier for the formation of the triple-C bond is not surmountable even in presence of the catalysts.
        
        For Sample A the production of alkanes and alkenes only start at higher temperatures than the other samples (at least 400\degrees). Moreover, the percentage of the compounds formed are significantly smaller than in the presence of the meteoritical samples, being particularly evident for the most abundant products of the reaction (\methane\ and \ethane). These findings clearly evidence the negligible activity of \sic\ and suggests that the activity detected is caused by the reaction occurring in the meteoritical samples.
        
        Sample E shows no production of any of the aforementioned products, indicating that no desorption or outgassing of organics from the meteorite is detected under these conditions.

    \subsection{Alcohols production}
    
\begin{figure}[h!]
        \begin{tabular}{m{8.5cm}}\includegraphics[width=9cm]{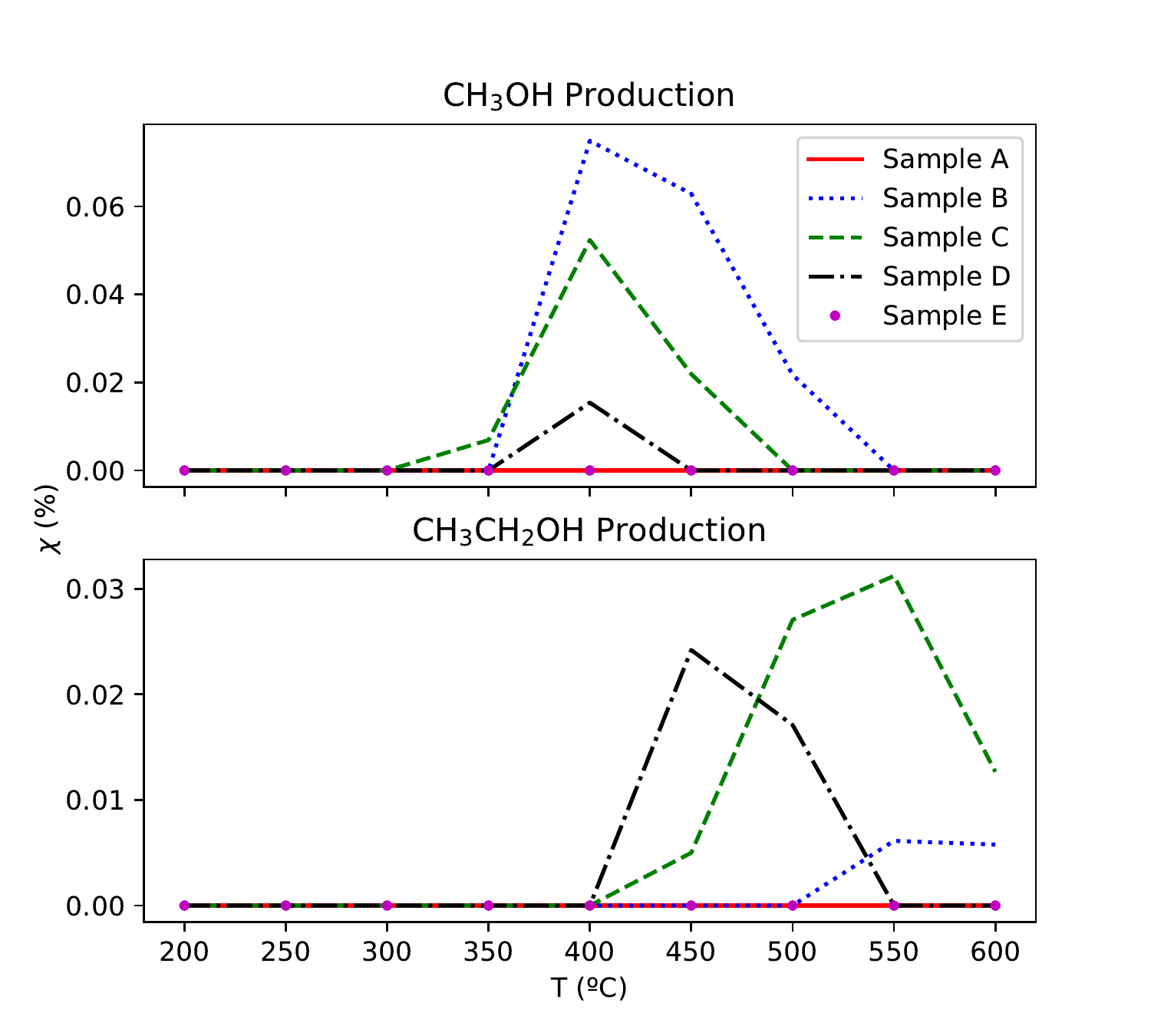} \\
        \end{tabular}
        \caption{Evolution of alcohols production with temperature. \methanol\ production (top) and \ethanol\ production (bottom).}
        \label{fig:alcohols_production}
        \end{figure}

        Figure \ref{fig:alcohols_production} shows the production of alcohols during the experiments. The profiles are rather different than those for hydrocarbon production. Formation of \methanol\ starts at 350\degrees, peaks at 400\degrees\ for all the samples and it stops before reaching 550\degrees, depending on the sample. Production of larger alcohols, mainly ethanol (\ethanol), is restricted to higher temperatures, starting at 400\degrees. The percentage of \ethanol\ starts to increase when \methanol\ starts decreasing, indicating a sequential alcohol production, (i.e., \ethanol\ can only form when the concentration of \methanol\ is large enough). The maximum production is at 550 \degrees\ for all samples except Sample D, which occurs at lower temperatures (450\degrees).
        
        Some larger alcohols such as propanol (\propanol) and butanol (\butanol) are detected, but only in a qualitative way. This indicates that the production of alcohols is not associated with the previous production of hydrocarbons so that both products are formed independently. No alcohols with an alkene chain have been detected, which sustains this hypothesis.
        
        Production of alcohols does not take place neither in Sample A nor Sample E, suggesting again that the products detected are being formed in the surface of the meteoritical material and are not coming from the activity of \sic\ or the meteorites themselves.
    
    \subsection{\cod\ production and carbon deposition}
    
        \begin{figure}[h!]
        \begin{tabular}{m{8.5cm}}\includegraphics[width=9cm]{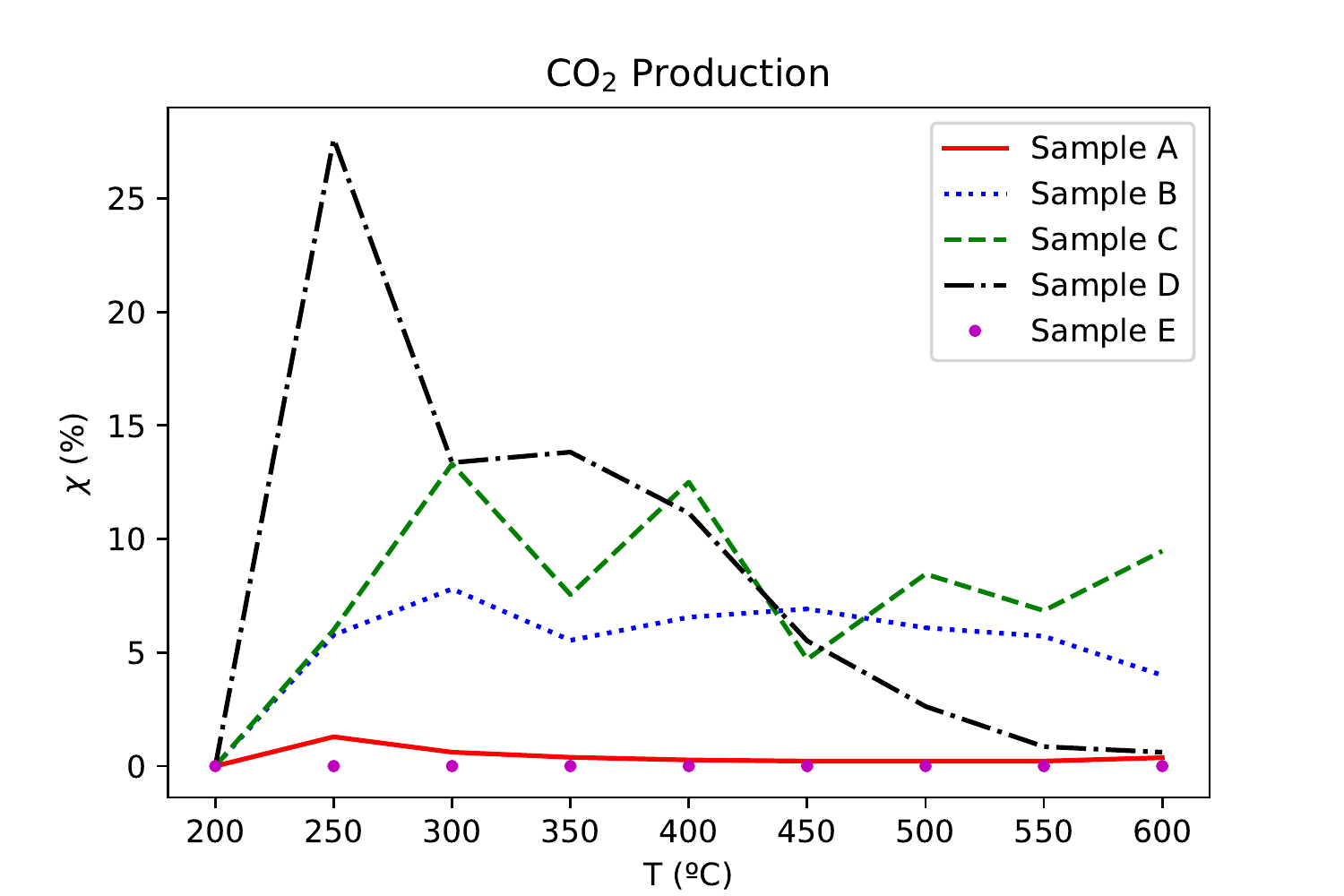} \\
        \end{tabular}
        \caption{Evolution of \cod\ production with temperature.}
        \label{fig:co2_production}
        \end{figure}
    
        \cod\ is continuously produced during the experiments, as shown in Figure \ref{fig:co2_production}. Its production starts at  250\degrees\ and it keeps relatively constant during all the experiment, the percentage of which depending on the sample. In view of that, \cod\ formation does not depend on the previous formation of hydrocarbons and/or alcohols. 
        Production of \cod\ is not detected for Sample E, suggesting again there is no outgassing from the meteorite matrix.
        
        The remaining percentage of the \co\ that has reacted but does not transform into products (and therefore, does not appear in the percentage counting) is deposited on the surfaces of the meteoritical samples as solid carbon, poisoning the surfaces, and, with time, likely preventing FTT reactions to take place in an efficient way (\citealt{Llorca1998}). The evolution of the surface poisoning is out of the scope of this paper.

    \subsection{Other compounds}
    
        Other compounds were detected during the experiments, but their evolution was not quantified. These compounds are mainly formaldehyde (\formaldehyde) and acetone (\acetone). They are formed at less than 400 \degrees, most possibly as byproducts or intermediates of the FTT reactions.

\section{Analysis and discussion} 

    \subsection{Organic compound production on chondritic materials}
     
    In this work, it is proven that chondritic materials can support FTT reactions. First of all, it can be concluded that the organics detected do not come from the meteoritic samples. The non-detection of any products on the blank (Sample E) proves that the detected products are formed by the interaction of the reactants with the meteorite surfaces and not coming from the desorption or the outgassing of the meteorite matrix. We note other reasons that suppoort that reactions are taking place in the reactor: i) the progressive decrease of the amount of \co\ detected in the GC with respect to the initial \co\ introduced in the reactor suggests that the products are actually formed from the reactants and hence not coming from the meteorites themselves, ii) the agreement of the distribution of products with the one expected from FTT reactions, the fact that no other species are detected, and the consecutive formation of products also indicate that the reactions are taking place on the surface of the meteorites, and iii) if the products were coming from the samples, their production will not be detected over the extended time of the experiments. It should also be noted that while CCs do contain a very small amount of organic matter, and could have been a possible source of contamination, OCs do not present any detectable percentage of organic material, accordingly those detected here are being formed by processes occurring in the reactor.
    
    It has also been checked that the reaction is not taking place because of the presence of \sic, by doing a blank analysis (Sample A). Those results show that the formation of hydrocarbons is negligible in comparison with the samples that present meteoritic material, especially for the most abundant products \methane\ and \ethane, and it only occurs at larger temperatures of at least 400 \degrees. Additionally, they do not show formation of alcohols, which is observed in the meteoritic samples. Overall, it is clear that the detected products are coming from reactions occurring inside the reactor and not from the samples itself, and that those reactions do not occur in the gas phase.
    
    The total conversion of \co\ in products with respect to solid carbon is relatively low, less than 25\%, but it is large enough to have quantitative results. In general, above 300 \degrees\ the production of alkanes and alkenes is relatively important, with selectivities between 25 and 75 \%\ of the total product conversion (without accounting for solid carbon). \cod\ is also produced in large amounts, with a selectivity that goes down from 100 to 25 \% as the temperatures rises. Alcohols are also produced but in minor  quantities, with selectivities not larger than 0.6 \%. Nevertheless, their formation is very important because it demonstrates the incorporation of O in an \oxygent-free environments, thereby forming polar groups that can result in molecules (i.e., \butanol\ and larger alcohols) of great importance for prebiotic chemistry and biochemistry (\citealt{Walde2005}).
    
    This work allowed to distinguish differences between the two classes of chondrites (samples B and D). Production of hydrocarbons starts at a slightly lower temperature for CCs than for OCs, though the percentage of products is roughly the same. The production of alcohols is similar as well, although it should be noted that sample D produces a larger amount of \ethanol. The similarities observed are attributed to the similar percentage of metallic inclusions in both chondrite-types, while the differences are probably caused by their different metallic composition and the different alteration degrees suffered by the materials.
    
    The activity of the metallic inclusions has been analyzed separately from the other phases (sample C). Results indicate that the overall production is larger than for samples B and D: it starts at lower temperatures and extends to larger temperatures. Some exceptions are noted: i) Production of \ethilene\ is lower for Sample C than for Sample D (NWA 801), and ii) \methanol\ production is low in comparison to the other samples. We attribute both events to a possible different composition of the metallic phases, which can favor the formation of some products in detriment of others, i.e. formation of \ethilene\ is not as favorable as \ethane\ for the metals present in Sample C, while formation of \methanol\ is reduced in favor of a larger and faster formation of \ethanol. Although in this work is not possible to directly correlate and quantify the activity with the metal content due to the lack of prior characterization of the complete composition of the meteoritical samples, our findings suggest that the metallic phases are the responsible of most of the surface activity of the samples, in detriment of the silicate phase. Moreover, previous experiments carried out by \citet{Ferrante2000}, using silicate phase analogues alone, did not detect any FTT products, supporting the hypothesis that the metallic phase (Fe in their case) was the responsible of the chemical activity. 
    
    The results of the present work are in agreement with other works available in the literature. Under conditions resembling those of the solar nebula (pressure $\approx$10$^{-4}$ atm and \hydrogen:\co\ $>$ 100:1) only hydrocarbons were produced, but with longer chains (C$_{1}$-C$_{4}$) (\citealt{Llorca1998}). When using Fe-doped silica instead, \citet{Ferrante2000} experiments, carried under similar conditions to this work (pressure $\approx$ 1 bar and \hydrogen:\co = 2-100), produced a similar hydrocarbon distribution, finding \methane, \ethane\ and \ethilene\ as the most abundant products. However, they also conclude that main activity was occurring in the metallic phase. Other works also obtained more complex hydrocarbons under similar conditions (\citealt{Gilmour2002}), such as aromatic compounds, which have not been produced in this work.  It should be noted that none of these works reported the production of alcohols or other oxygenated compounds. 
    
    The differences between the different works can be explained as follows. First, the experimental conditions are different. Here it is considered an Earth-like environment, i.e., larger pressures and \hydrogen:\co\ = 4. As expected, larger pressures give faster reactions (reaction times of $\sim$3h, this work and \citealt{Ferrante2000} vs $>$ 10$^{2}$ h, \citealt{Llorca1998}). Moreover, our \hydrogen:\co\ ratio probably favors the formation of alcohols in detriment of hydrocarbons (\hydrogen:\co = 4, this work vs. \hydrogen:\co = 250, \citealt{Llorca1998}). Secondly, actual meteoritic samples are used while the cited works used synthetic metallic alloys. It is possible that alteration processes undergone by the meteoritic materials produced structural and electronic changes on their surfaces, giving rise to a different catalytic behavior. In addition, meteoritic materials do not only contain Fe in their composition, usually used as a metallic phase analogue, but also other metals, such as Ni or Co, which are known to have an effect on the selectivity and product distribution of FTT reactions (\citealt{Mahmoudi2017}). This could favor the formation of oxygen-bearing compounds, which are not reported in other works. In order to totally understand the effect of the different alteration degrees of the samples in the chemical outcome and the actual relationship between the metal quantity and the catalytic activity, more similar experiments are needed using a more extensive sampling of chondrite types, which suffered different degrees of alteration and contain other different amounts of metallic inclusions, such as CIs and CBs (which contain 0 \vol\ and 60-80 \vol\ abundance of metals respectively).

    \subsection{The early Earth atmosphere and problems of the Fischer-Tropsch mechanism}
    
    One of the main conundrums to assess if these reactions could have taken place in an early-Earth environment relies on the real atmospheric composition at that time. The work by Miller and Urey (\citealt{Miller1953}) assumed a strongly reducing atmosphere, composed by \hydrogen, \methane, \ammonia\ and \water, which is an ideal environment for FTT reactions. However, this vision has been questioned over the years. Based on geological arguments, the early atmosphere came from the degassing of Earth interior, containing mainly  \cod, \water\ and \nitrogen, which does not support the formation of \methane\ and \ammonia\ (\citealt{Zahnle2010}, \citealt{Lammer2018}, \citealt{Gronoff2020}).
    
    Nevertheless, additional information on the processes that happened in the primeval Earth points to a more complex story. The first atmosphere probably contained gases trapped from the solar nebula during accretion, mainly \hydrogen, \co, \methane\ and \ammonia. In small planets, such as the Earth, when the protostellar envelope vanishes the first atmosphere is likely to escape and outgassing of the planet forms the secondary atmosphere. The large amount of impacts during the formation of Earth could have helped the emergence of this atmosphere, since the volatilization of parts of the silicates of the mantle could have allowed the presence of  volatile elements, e.g. \co, in the atmosphere for a relatively long time (\citealt{Lammer2018}). 
    
    A consequence of these processes is that, even that \co-containing atmospheres could not be stable, they probably existed as steady states for periods long enough so \co\ could react with \hydrogen\, triggering FTT processes, as well as other fundamental reactions such as the Haber-Bosch (HB) synthesis (i.e., ammonia formation from \nitrogen\ and \hydrogen), thereby giving rise to a richer and more complex chemistry in the planet surface. These processes are envisioned to happen on vapor plumes produced after the impact of meteorites on Earth surface (\citealt{Trigo2013}, \citealt{Lammer2018}), when reducing gases were locally released and pressure and temperature were high enough.

\section{Summary}

In this work, FTT reactions have been investigated in the presence of different meteoritical samples and under conditions resembling those of the early Earth. These laboratory experiments demonstrate that the undifferentiated meteorites can act as a surface support to promote the synthesis of different organic compounds from \hydrogen\ and \co\ mixtures. The following conclusions have been reached:

         \begin{itemize}
         
            \item FTT reactions can occur under planetary conditions with the surface support of chondritic materials.
         
            \item FTT reactions under these particular conditions yield the synthesis of hydrocarbons (\methane, \ethane\ and \ethilene) as well as alcohols (\methanol\ and \ethanol) and other oxygenated compounds, such as \formaldehyde\ and \acetone\ even in \oxygent-free reaction conditions. A large amount of \cod\ is also produced.
            
            \item The detection of products is associated with reactions happening on the surfaces of the meteorites and not due to the desorption of organic contents already present in the materials themselves.The activity is mainly associated with the metallic phases, since they show larger productivity than other mineral phases (e.g., silicates). 
            
            \item Future work is needed to fully characterize the possible catalytic properties of different chondrite-types, with special focus on the influence of the different percentages of metal inclusions and different alteration degrees, and to determine if these materials can also promote other reaction-types crucial for prebiotic chemistry, such as the HB synthesis and further \nitrogen\ chemistry or sulfur-/phosphorous-based chemistry.
            
        \end{itemize}

\section*{Acknowledgments}

J.M.T.-R. acknowledges financial support from the Spanish Ministry (PGC2018-097374-B-I00). J.L. is grateful to ICREA Academia program and funding from Generalitat de Catalunya (2017 SGR 128). AR is indebted to the “Ramón  y  Cajal”  program, MINECO project  CTQ2017-89132-P,  DIUE  project  2017SGR1323, and funding from the European Research Council (ERC) under the European Union’s Horizon 2020 research and innovation programme (grant agreement No. 865657) for the project “Quantum Chemistry on Interstellar Grains” (QUANTUMGRAIN).

\vspace{-10pt}
\bibliographystyle{aa}
\bibliography{mybiblio}

\clearpage
\onecolumn
\begin{appendix}
    \section{Experimental chromatograms}
    
\begin{figure}[h!]
            \centering
            \includegraphics[scale=1]{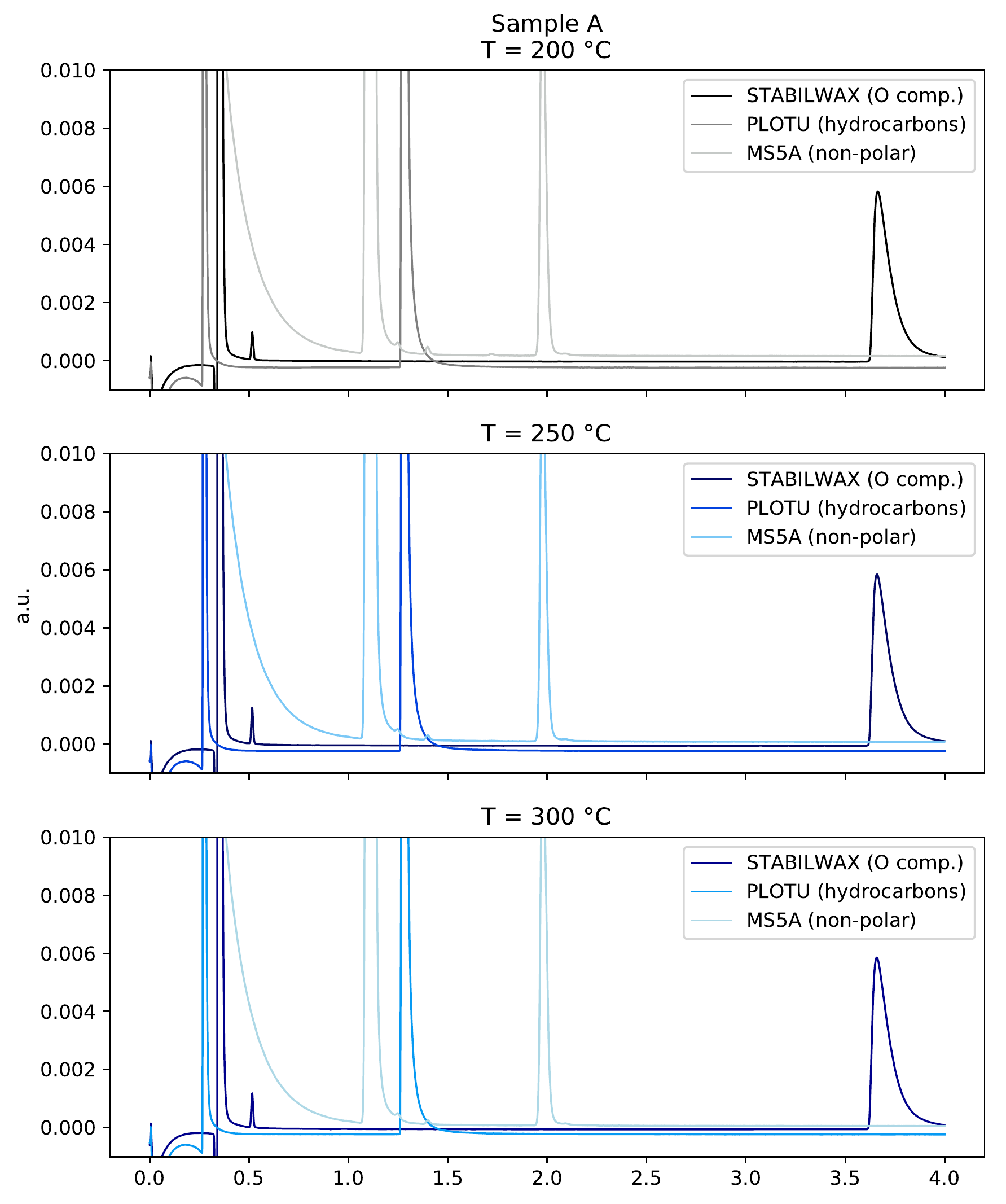}
            \caption{Normalized chromatograms at 200, 250 and 300 \degrees C for Sample A.}
            \label{fig:Chrom_A1}
        \end{figure}

\begin{figure}[h!]
            \centering
            \includegraphics[scale=1]{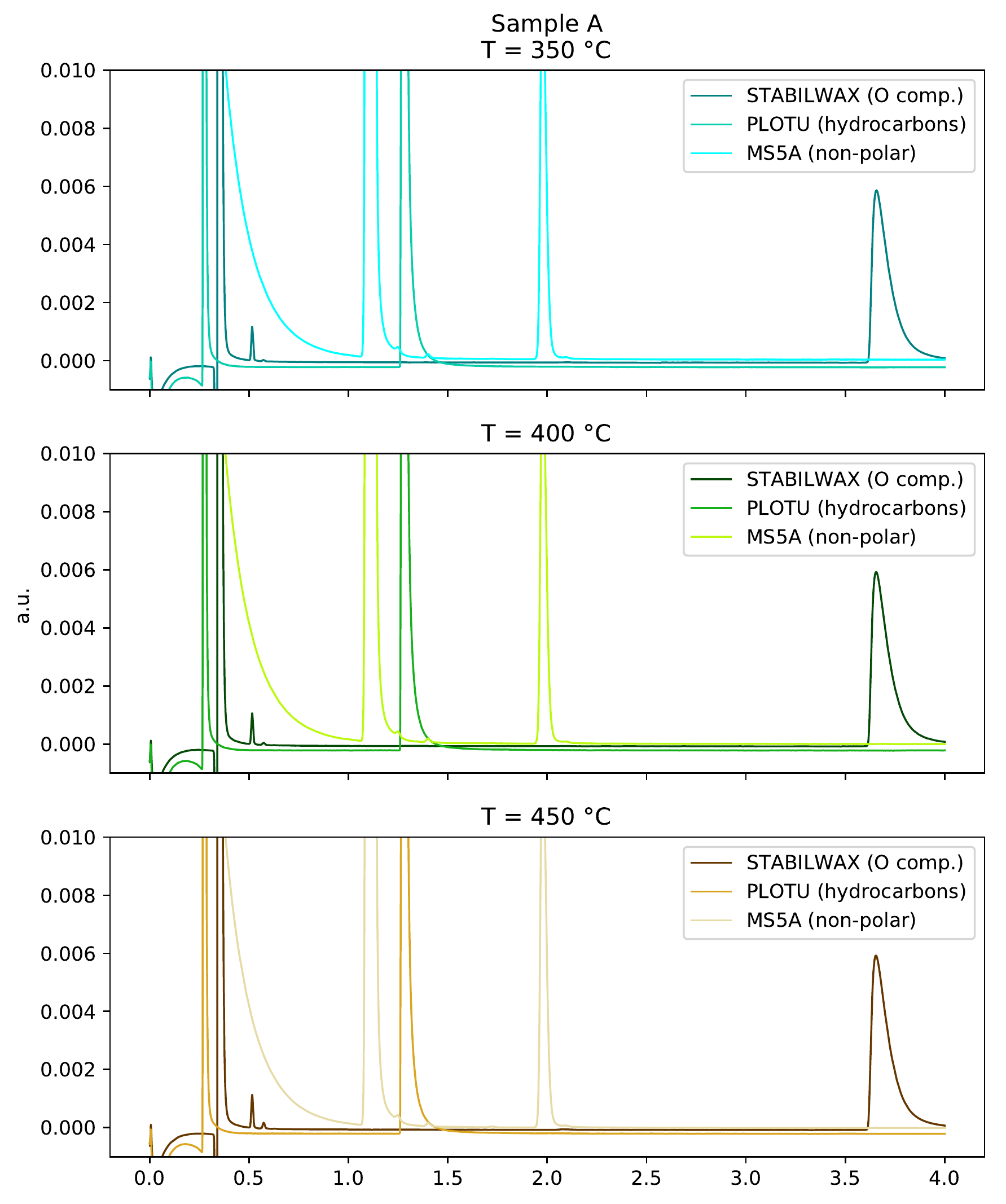}
            \caption{Normalized chromatograms at 350, 400 and 450 \degrees C for Sample A.}
            \label{fig:Chrom_A2}
        \end{figure}

\begin{figure}[h!]
            \centering
            \includegraphics[scale=1]{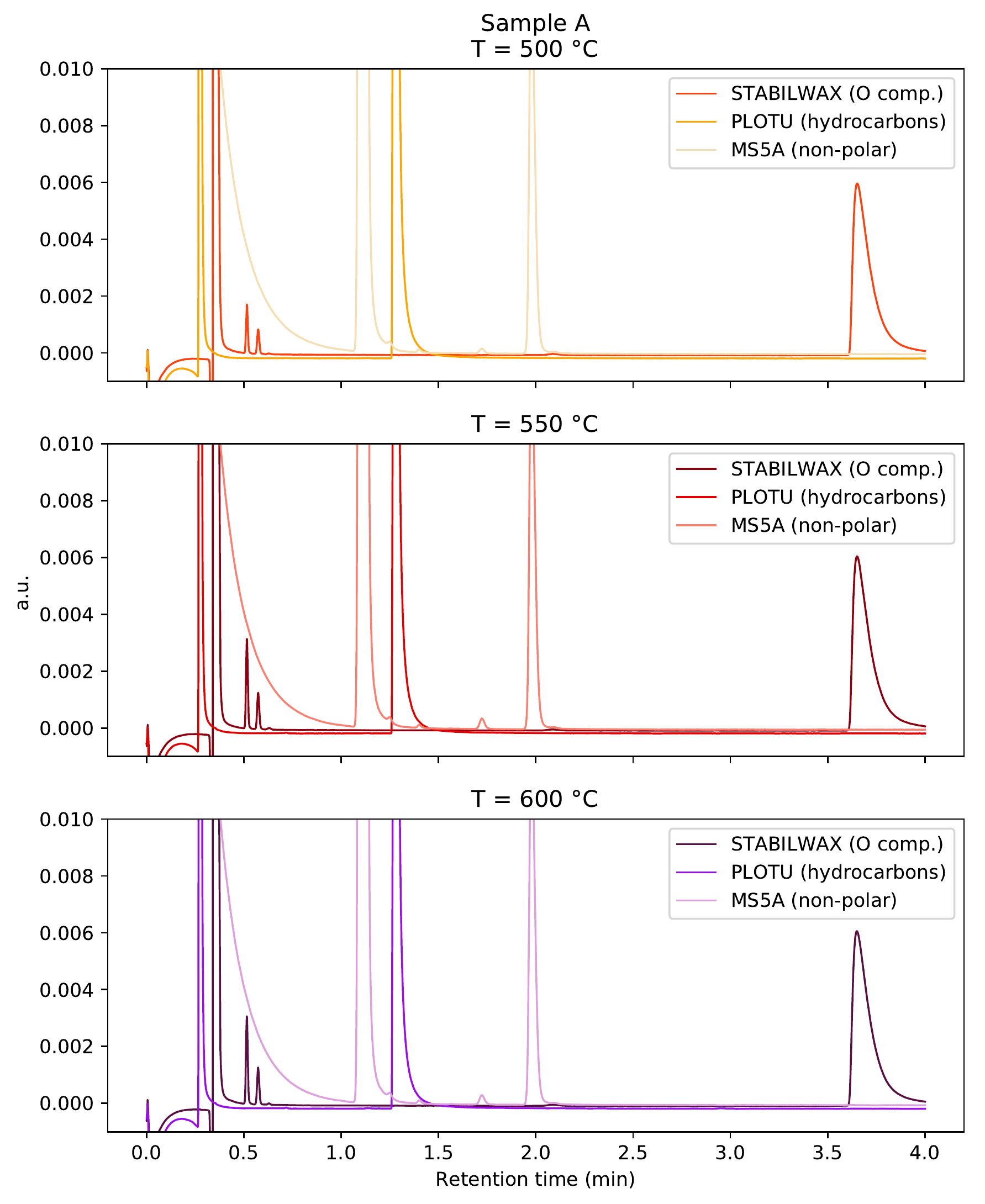}
            \caption{Normalized chromatograms at 500, 550 and 600 \degrees C for Sample A.}
            \label{fig:Chrom_A3}
        \end{figure}

\begin{figure}[h!]
            \centering
            \includegraphics[scale=1]{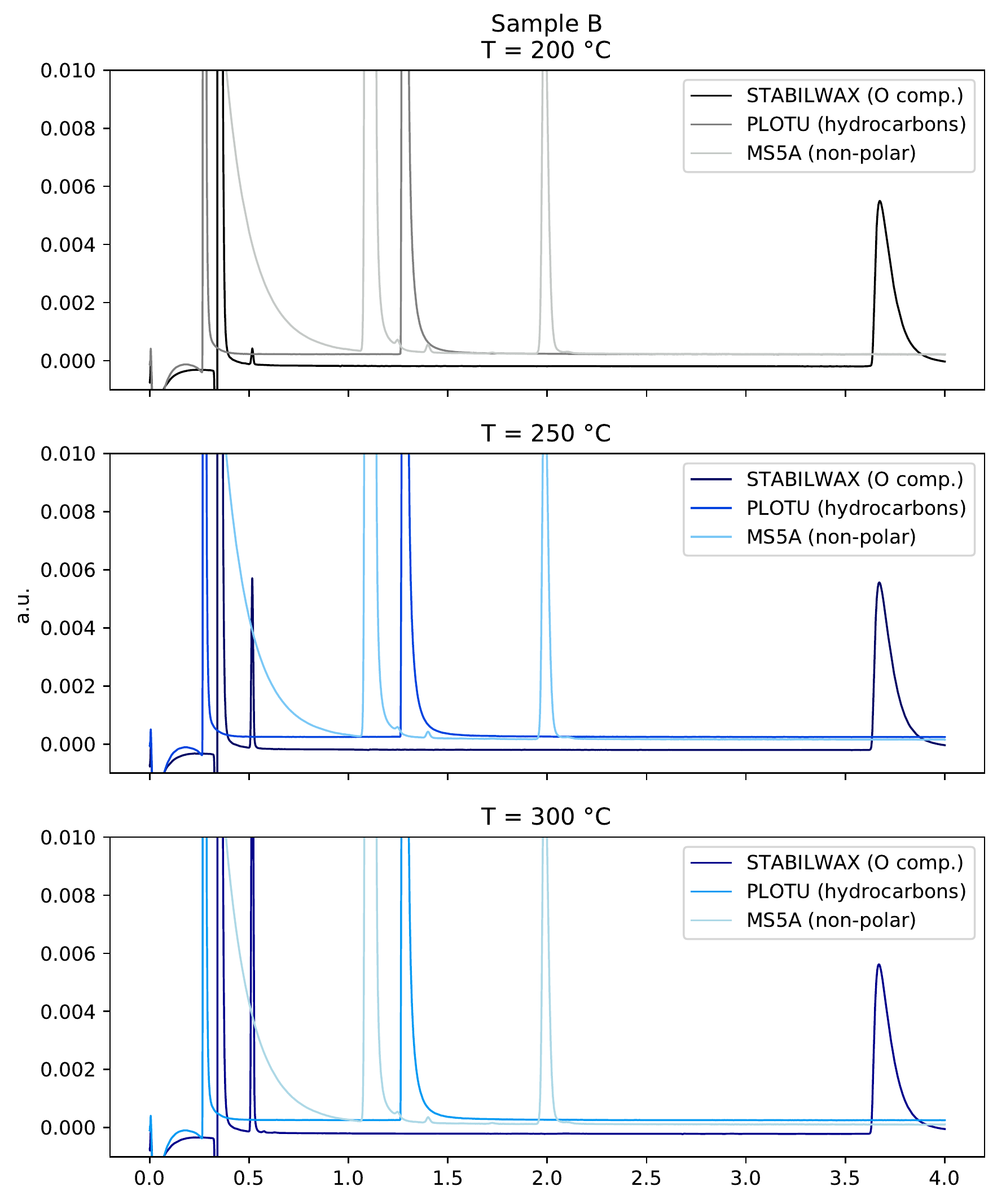}
            \caption{Normalized chromatograms at 200, 250 and 300 \degrees C for Sample B.}
            \label{fig:Chrom_B1}
        \end{figure}

\begin{figure}[h!]
            \centering
            \includegraphics[scale=1]{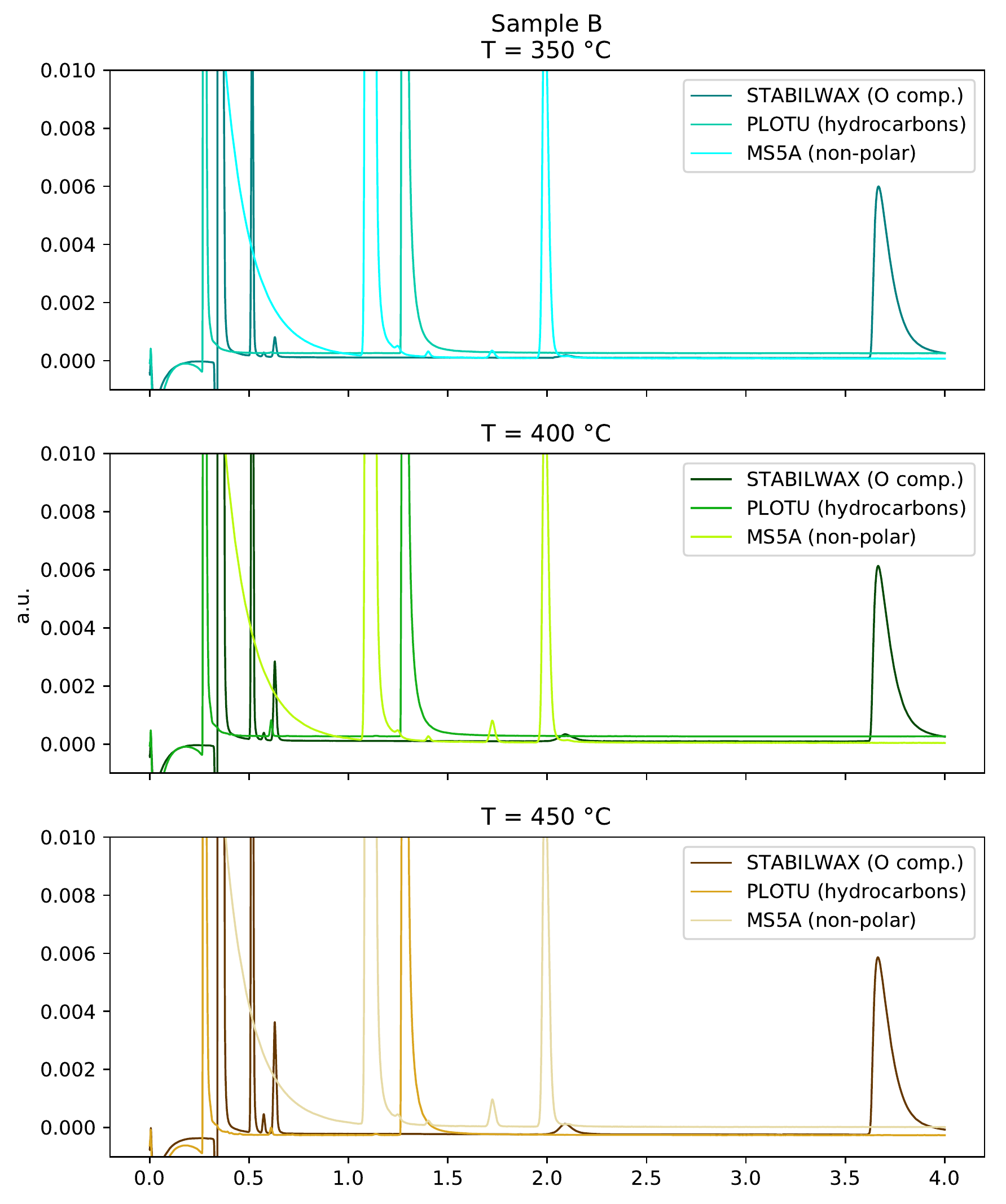}
            \caption{Normalized chromatograms at 350, 400 and 450 \degrees C for Sample B.}
            \label{fig:Chrom_B2}
        \end{figure}

\begin{figure}[h!]
            \centering
            \includegraphics[scale=1]{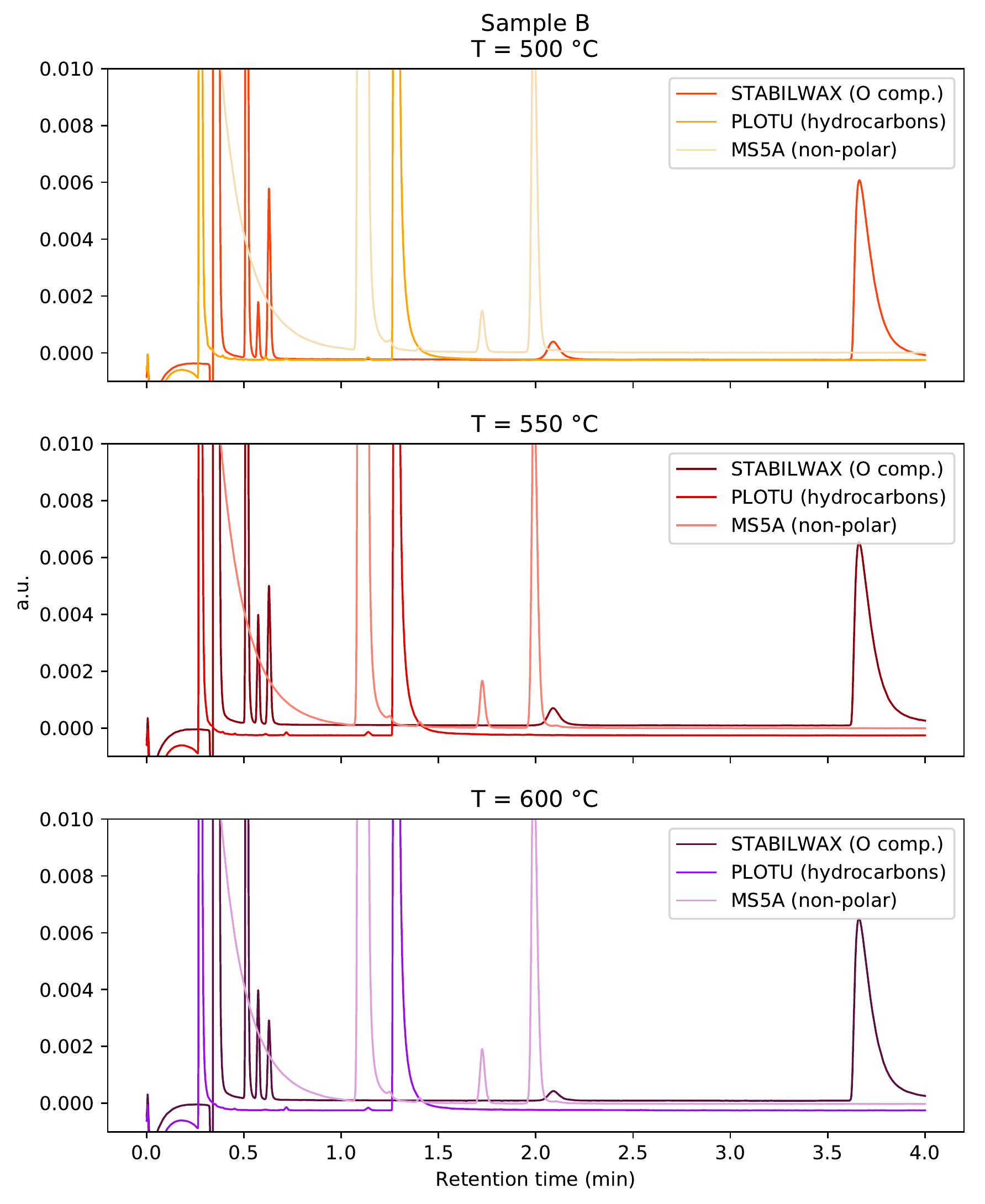}
            \caption{Normalized chromatograms at 500, 550 and 600 \degrees C for Sample B.}
            \label{fig:Chrom_B3}
        \end{figure}

\begin{figure}[h!]
            \centering
            \includegraphics[scale=1]{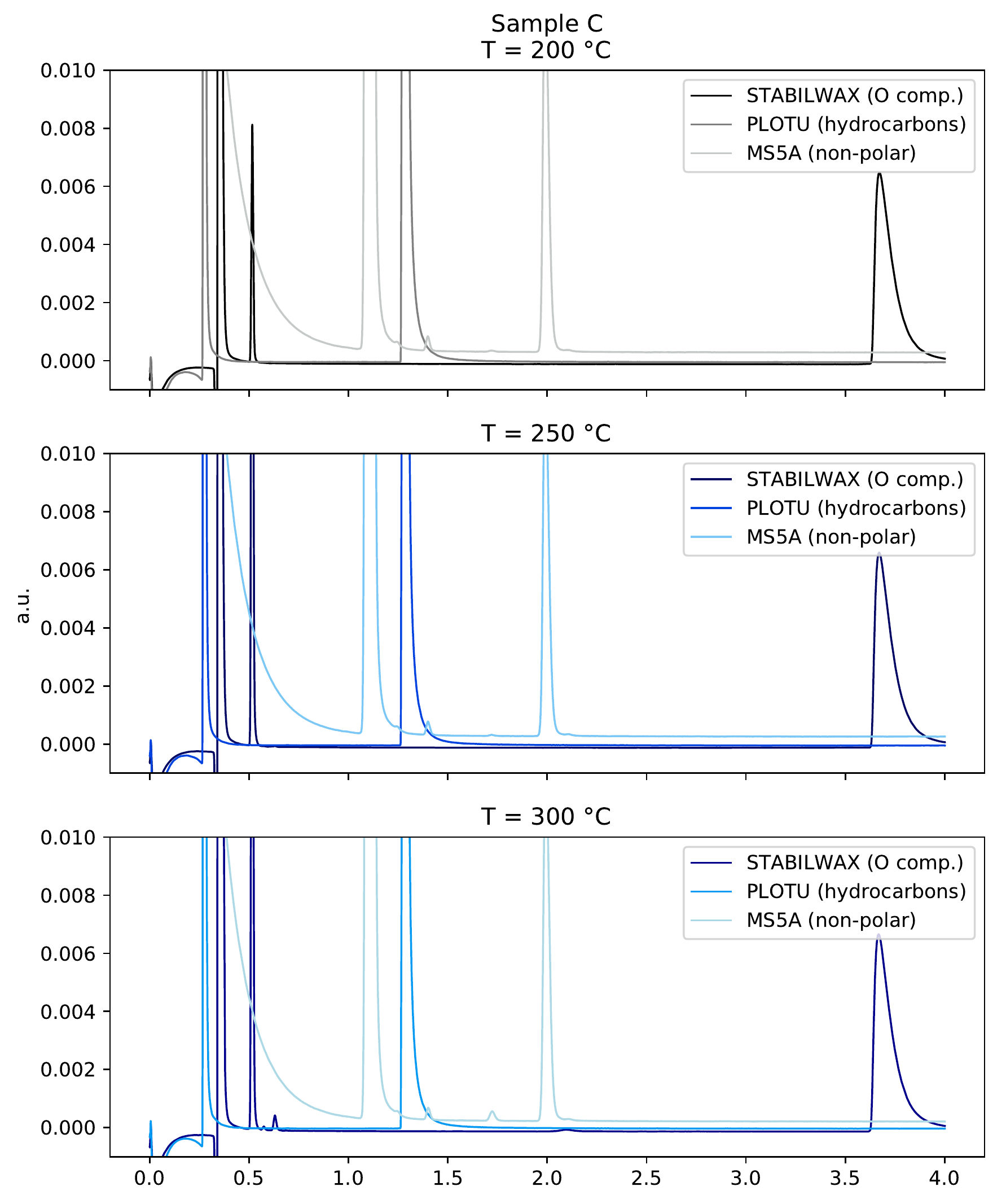}
            \caption{Normalized chromatograms at 200, 250 and 300 \degrees C for Sample C.}
            \label{fig:Chrom_C1}
        \end{figure}

\begin{figure}[h!]
            \centering
            \includegraphics[scale=1]{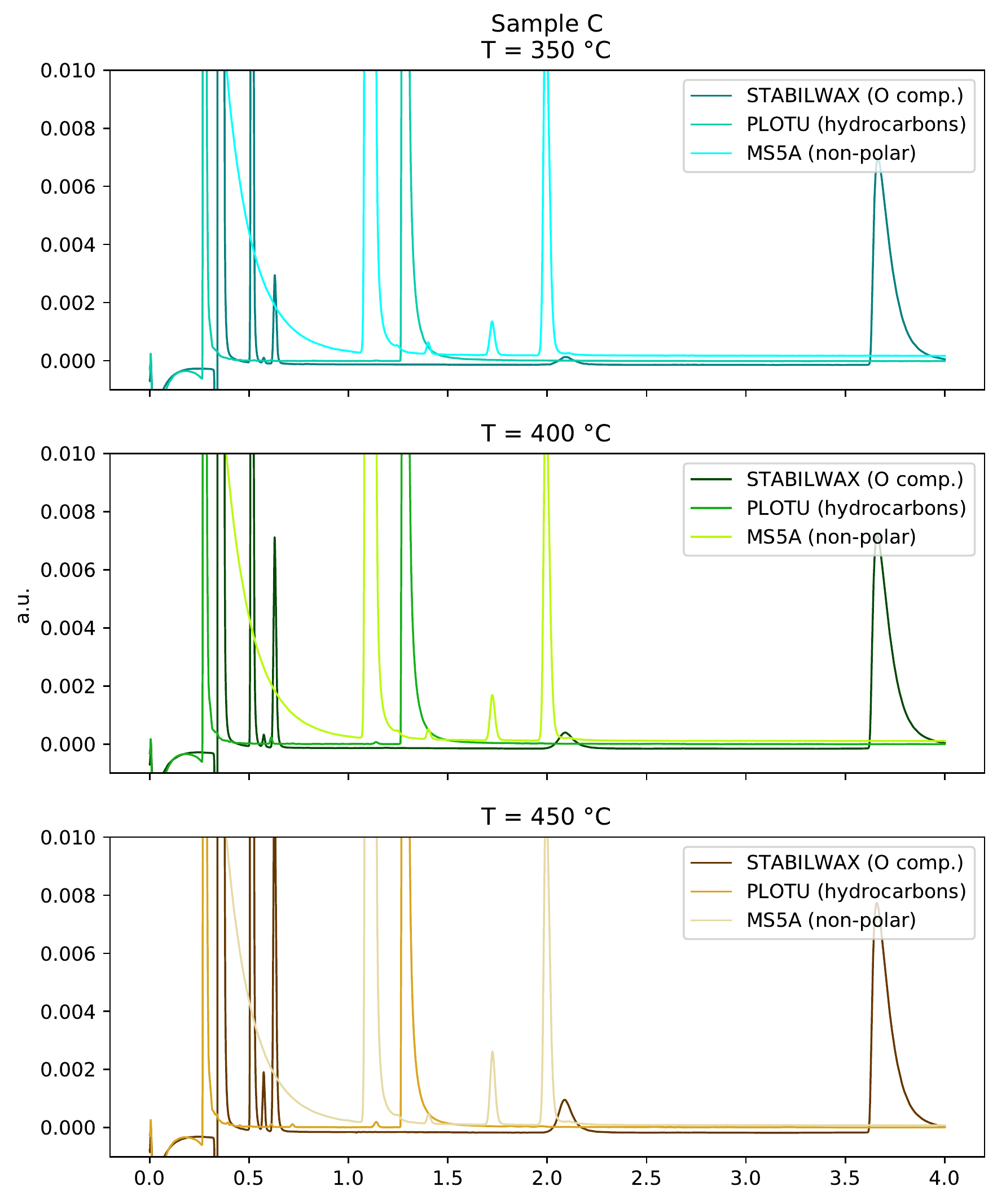}
            \caption{Normalized chromatograms at 350, 400 and 450 \degrees C for Sample C.}
            \label{fig:Chrom_C2}
        \end{figure}

\begin{figure}[h!]
            \centering
            \includegraphics[scale=1]{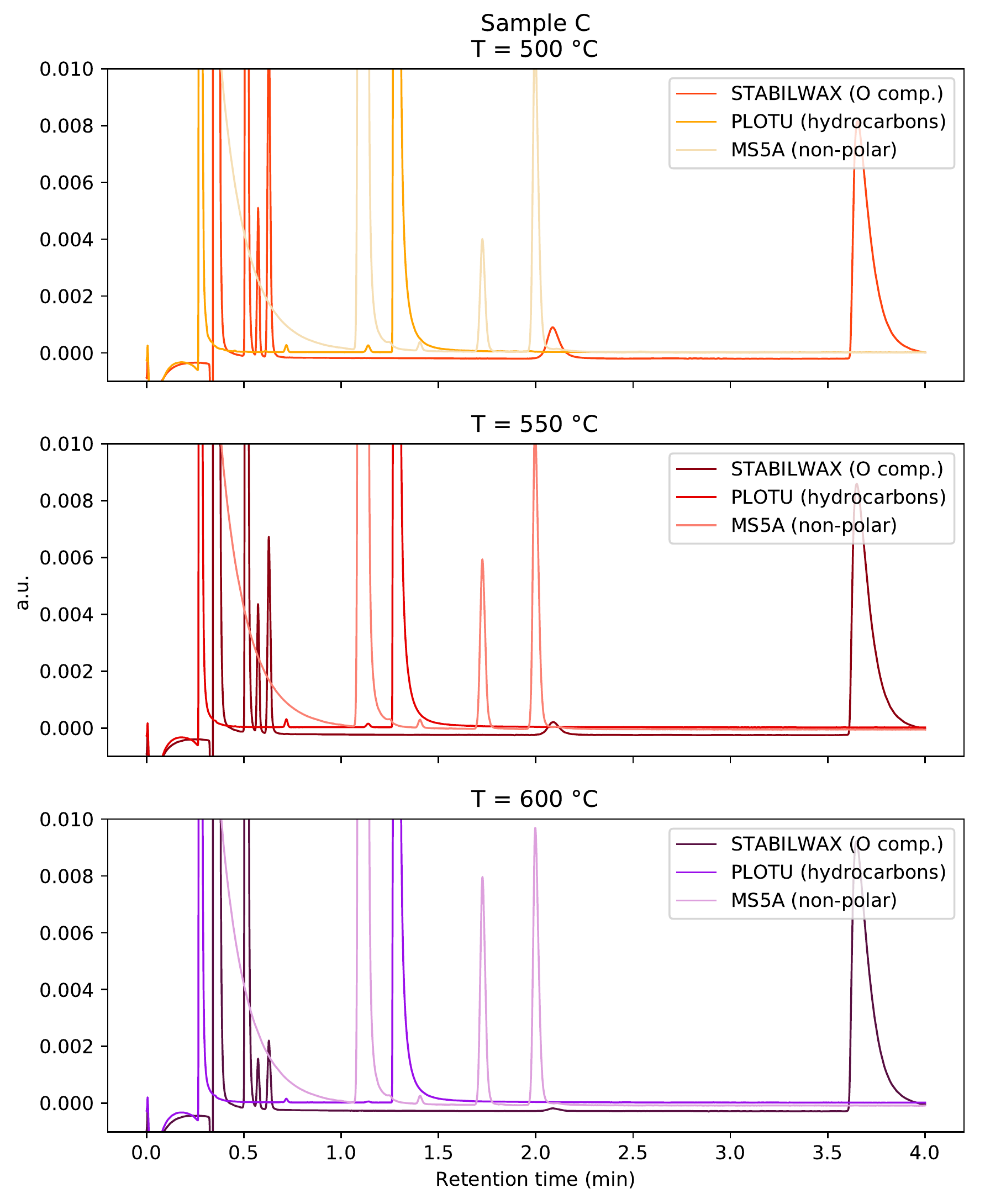}
            \caption{Normalized chromatograms at 500, 550 and 600 \degrees C for Sample C.}
            \label{fig:Chrom_C3}
        \end{figure}

\begin{figure}[h!]
            \centering
            \includegraphics[scale=1]{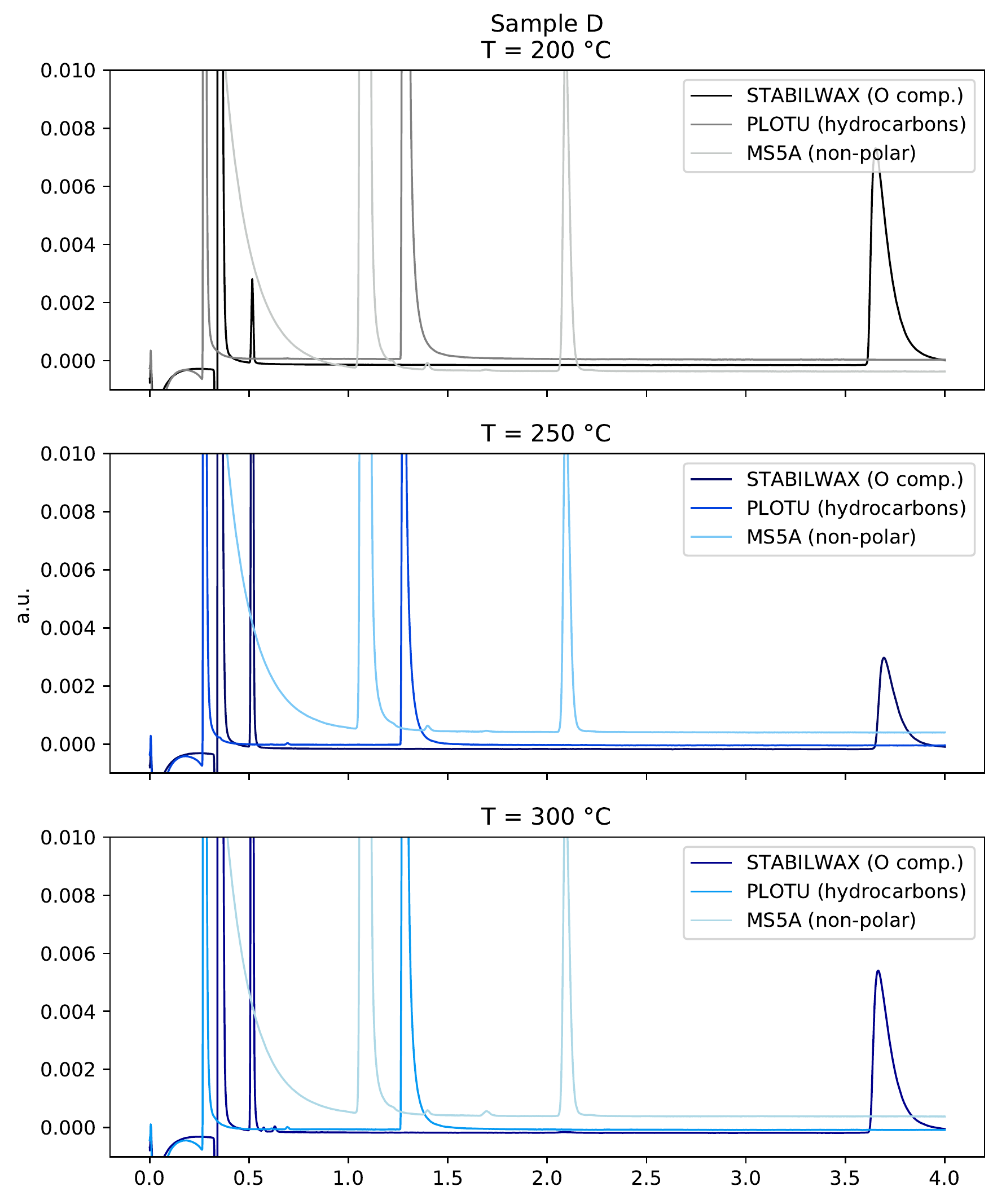}
            \caption{Normalized chromatograms at 200, 250 and 300 \degrees C for Sample D.}
            \label{fig:Chrom_D1}
        \end{figure}

\begin{figure}[h!]
            \centering
            \includegraphics[scale=1]{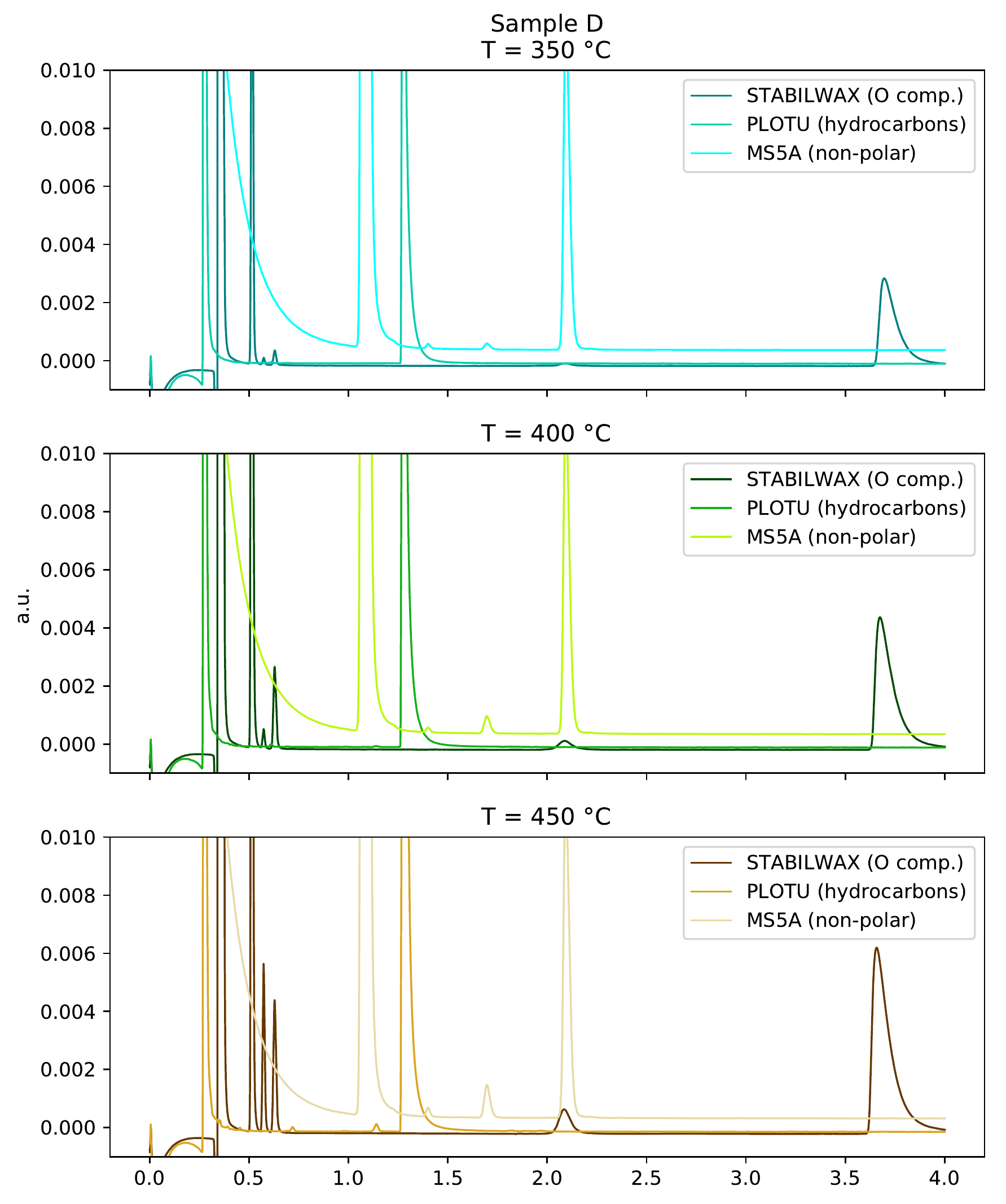}
            \caption{Normalized chromatograms at 350, 400 and 450 \degrees C for Sample D.}
            \label{fig:Chrom_D2}
        \end{figure}

\begin{figure}[h!]
            \centering
            \includegraphics[scale=1]{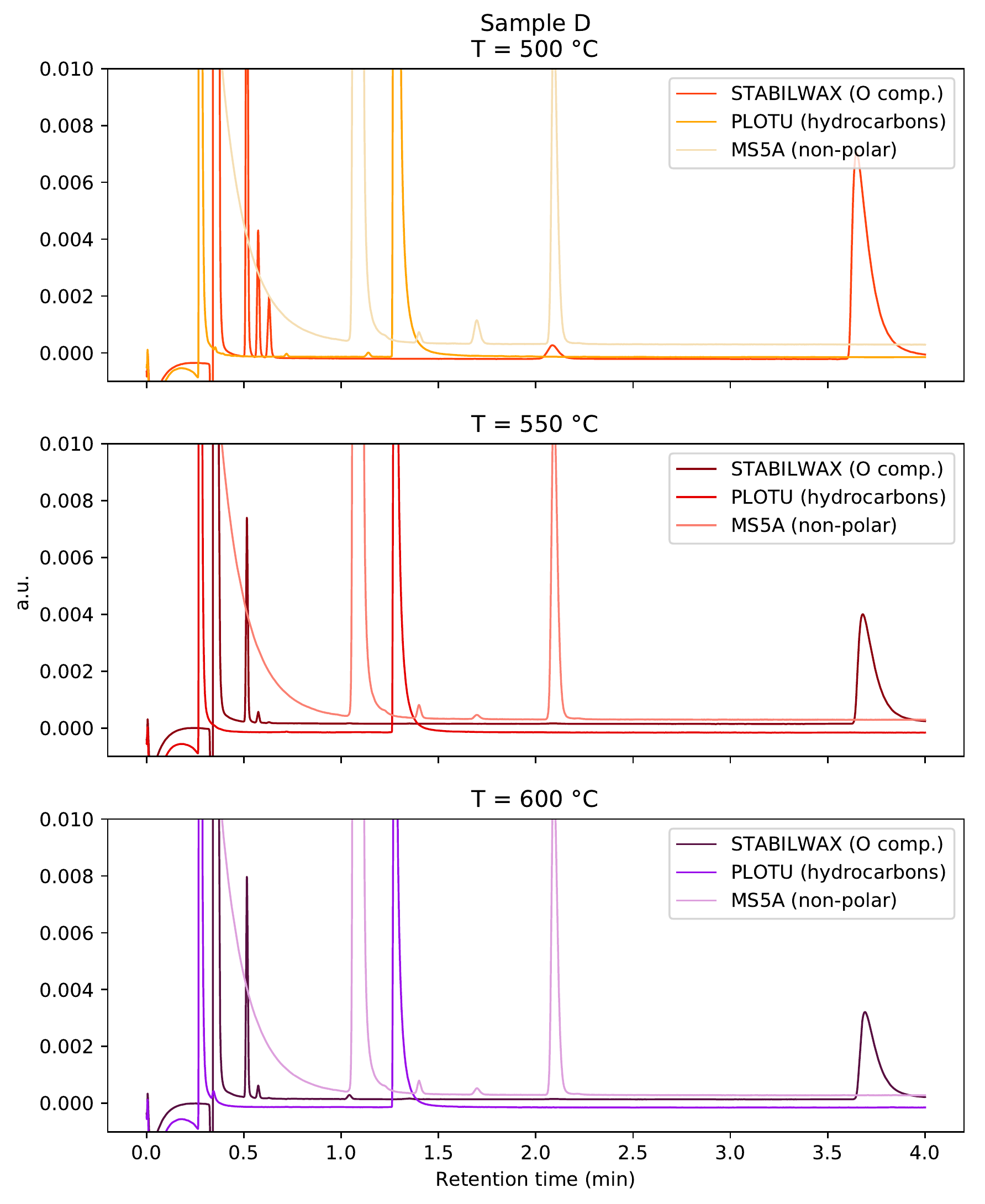}
            \caption{Normalized chromatograms at 500, 550 and 600 \degrees C for Sample D.}
            \label{fig:Chrom_D3}
        \end{figure}

\begin{figure}[h!]
            \centering
            \includegraphics[scale=1]{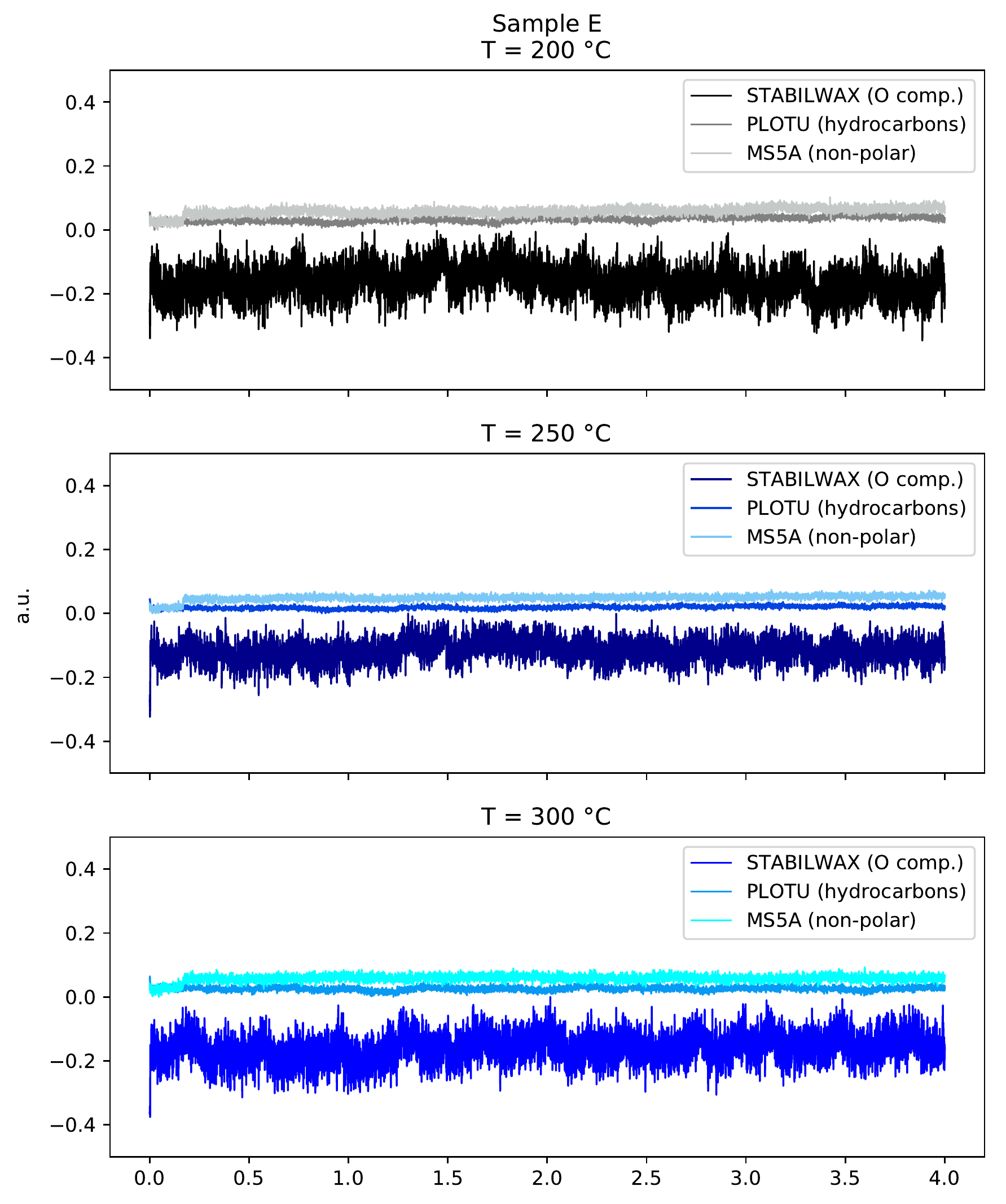}
            \caption{Normalized chromatograms at 200, 250 and 300 \degrees C for Sample E. The plot shows only noise, due to the lack of any reactant gas or product detected.}
            \label{fig:Chrom_E1}
        \end{figure}

\begin{figure}[h!]
            \centering
            \includegraphics[scale=1]{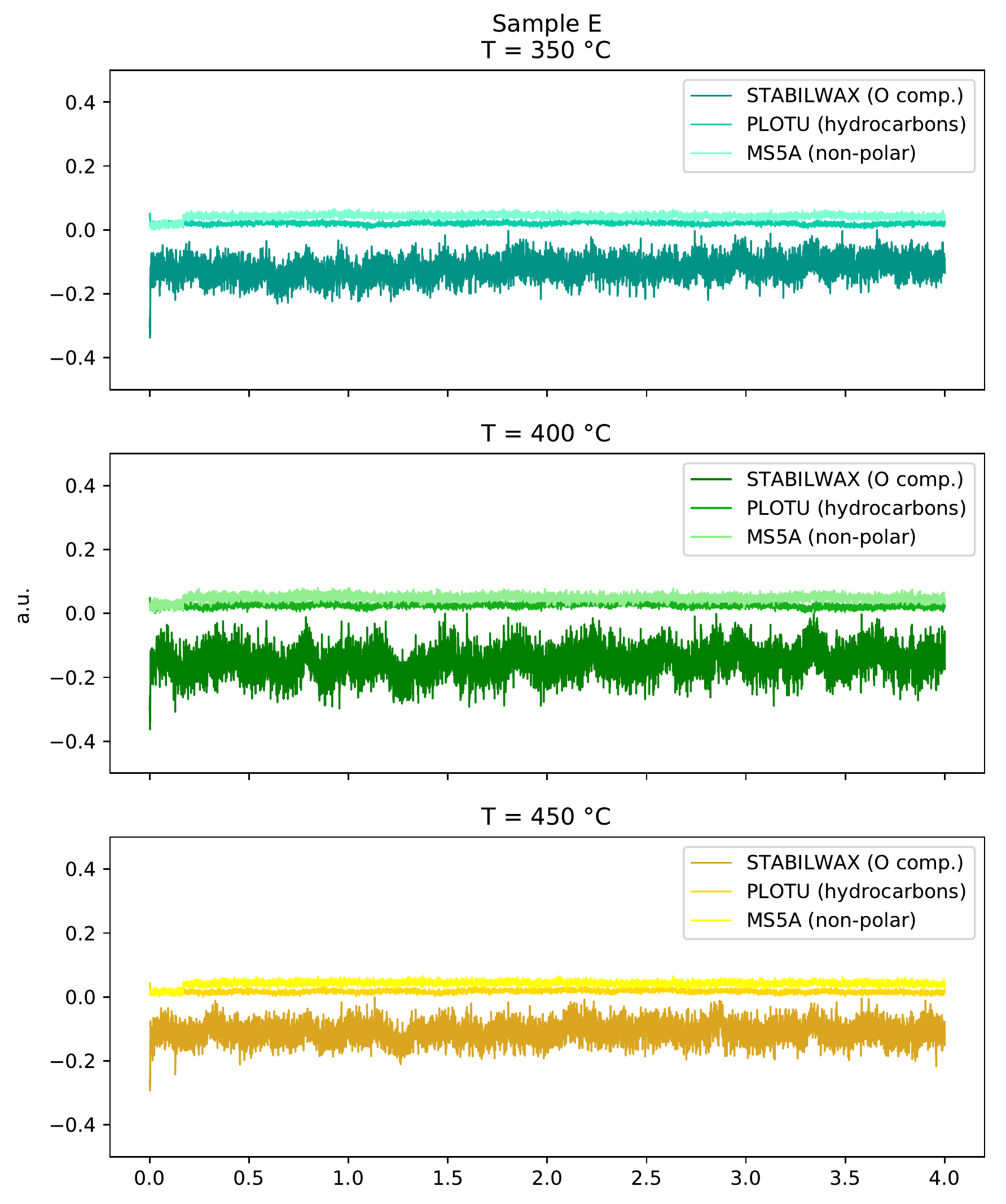}
            \caption{Normalized chromatograms at 350, 400 and 450 \degrees C for Sample E. The plot shows only noise, due to the lack of any reactant gas or product detected.}
            \label{fig:Chrom_E2}
        \end{figure}

\begin{figure}[h!]
            \centering
            \includegraphics[scale=1]{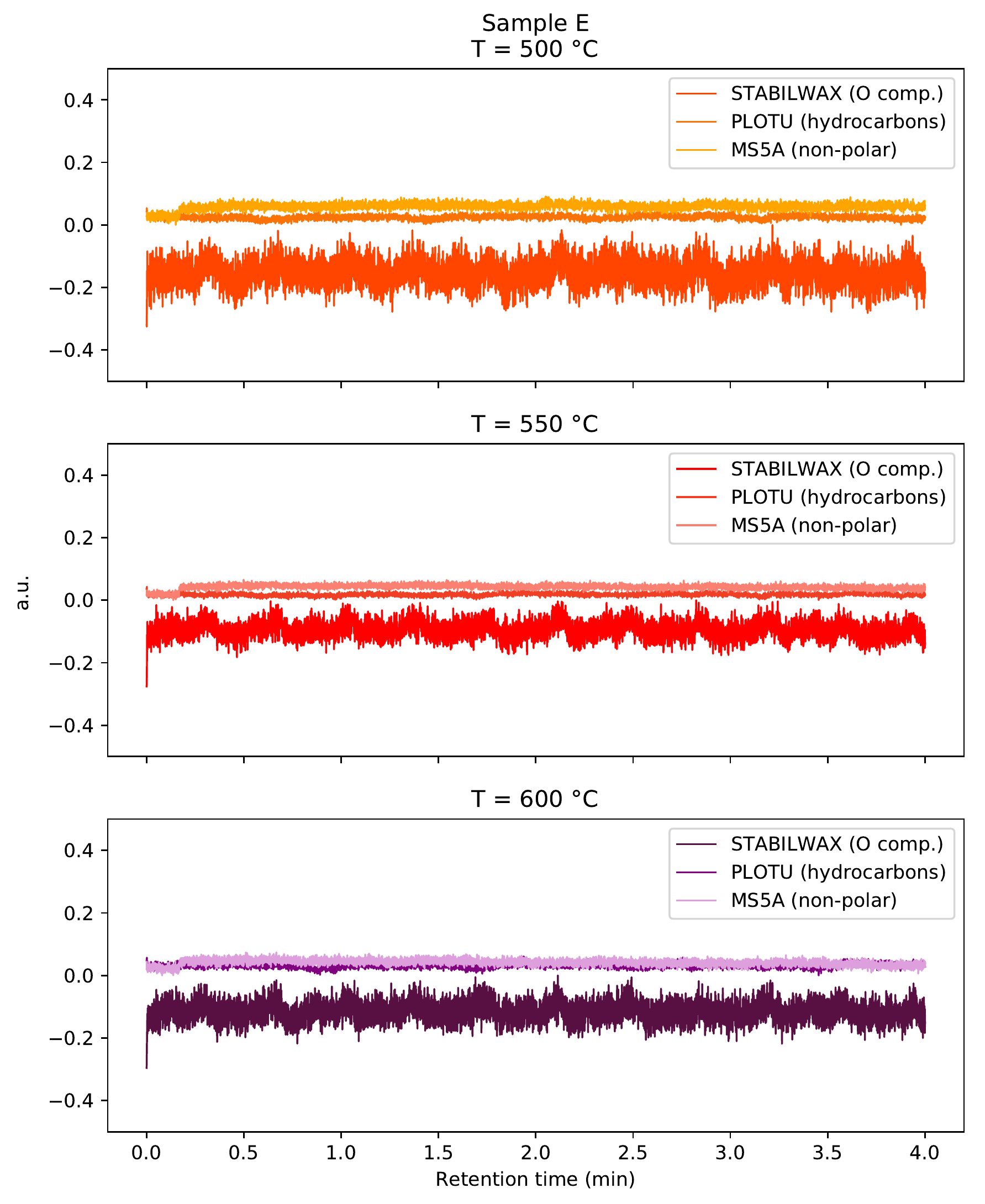}
            \caption{Normalized chromatograms at 500, 550 and 600 \degrees C for Sample E. The plot shows only noise, due to the lack of any reactant gas or product detected.}
            \label{fig:Chrom_E3}
        \end{figure}

        Figures \ref{fig:Chrom_A1} to \ref{fig:Chrom_E3} show the chromatograms of each GC column, for each sample analyzed and at each reference temperature. Table \ref{tab:retention_times} shows a guidance to the retention times used to assign the detected peaks. Note that the different GC columns have different baselines levels, but each analysis has been realized with respect to the correspondent level, and therefore there should not be any baseline effects in the computed product percentage. 
    
        \begin{table}[h!]
            \centering
            \begin{tabular}{c c | c c | c c}
                \hline
                \multicolumn{2}{c}{STABILWAX} & \multicolumn{2}{|c|}{PLOTU} &  \multicolumn{2}{c}{MS5A} \\
                \hline
                Molecule & Ret. time (min) & Molecule & Ret. time (min) & Molecule & Ret. time (min) \\
                \hline
                \formaldehyde & 0.55 & \ethilene & 0.68 & \hydrogen & 1.15 \\
                \acetone & 0.65 & \ethane & 0.75 & \nitrogen & 1.25 \\
                \methanol & 0.70 & \co & 1.35  & \methane & 1.75 \\
                \ethanol & 2.10 & & & \cod & 2.00\\
            \end{tabular}
            \caption{Guidance retention time for all the molecules detected by each GC column.}
            \label{tab:retention_times}
        \end{table}

        \clearpage
        
        \clearpage
        
        \clearpage
        
        \clearpage
        
        \clearpage
        
        \clearpage
        
        \clearpage
        
        \clearpage
        
        \clearpage
        
        \clearpage
        
        \clearpage
        
        \clearpage
        
        \clearpage
        
        \clearpage
        
\end{appendix}

\end{document}